\documentclass[journal]{IEEEtran}
\usepackage{tikz}
\usetikzlibrary{decorations.pathmorphing}
\usepackage[intlimits]{amsmath}
\usepackage{amsfonts}
\usepackage{amssymb}

\usepackage{booktabs}
\usepackage{multirow}

\usetikzlibrary{patterns}

\usepackage{algorithm}
\usepackage{algpseudocode}
\usepackage{setspace}
\usepackage{cite}
\newfloat{algorithm}{t}{lop}
\usepackage[normalem]{ulem}

\usepackage{capekCommands}
\graphicspath{figures/}

\newcommand{\ju}{\J}

\renewcommand{\Re}{\operatorname{Re}}	

\providecommand*{\unit}[1]{\ensuremath{\mathrm{\,#1}}}
\definecolor{breg}{rgb}{0.2,0.6,0.8}%
\definecolor{preg}{rgb}{0.8,0.2,0.2}%
\definecolor{reg}{RGB}{218,165,32}

\def\opC{0.7}

\newcommand{\cSub}{\T{c}} 
\newcommand{\bSub}{\T{b}} 
\newcommand{\reg}{\varOmega}
\newcommand{\regc}{\reg_\cSub} 
\newcommand{\regu}{\reg_\bSub} 
\providecommand*{\wtM}[1]{\widetilde{\mathbf#1}}

\providecommand{\nS}{\tilde{\M{S}}}

\renewcommand{\TBs}[1]{}
\renewcommand{\TB}[1]{\textcolor{black}{#1}}

\title{Theory and Computation of Substructure Characteristic Modes}
\author{
Mats Gustafsson, \IEEEmembership{Senior Member, IEEE}, 
Lukas Jelinek,
Miloslav Capek, \IEEEmembership{Senior Member, IEEE},\\
Johan Lundgren, \IEEEmembership{Member, IEEE}, 
and
Kurt Schab, \IEEEmembership{Member, IEEE} 
\thanks{Manuscript received \today; revised \today. This work was supported by ELLIIT - an Excellence Center at Linkoping-Lund in Information Technology and the Czech Science Foundation under project~\mbox{No.~21-19025M}.}
\thanks{M. Gustafsson and J. Lundgren are with Lund University, Lund, Sweden, (e-mails: \{mats.gustafsson, johan.lundgren\}@eit.lth.se).}
\thanks{L. Jelinek and M. Capek are with Czech Technical University in Prague, Czech Republic (e-mails: \{lukas.jelinek,miloslav.capek\}@fel.cvut.cz).}
\thanks{K. Schab is with the Santa Clara University, Santa Clara, USA (e-mail: kschab@scu.edu).}
\thanks{Color versions of one or more of the figures in this paper are available online at http://ieeexplore.ieee.org.}
}

\begin{document}

\maketitle

\begin{abstract}
The problem of substructure characteristic modes is \TB{developed} using a scattering matrix-based formulation, generalizing subregion characteristic mode decomposition to arbitrary computational tools. It is shown that the \TB{modes of the } scattering formulation \TB{are} identical to the \TB{modes of the } classical formulation based on the background Green's function for lossless systems \TB{under conditions where both formulations can be applied}. The scattering formulation, however, opens a variety of new subregion scenarios unavailable within previous formulations, including cases with lumped or wave ports or subregions in circuits. Thanks to its scattering nature, the formulation is solver-agnostic with the possibility to utilize an arbitrary full-wave method.
\end{abstract}


\section{Introduction}
Characteristic mode decomposition~\cite{Montgomery+Dicke+Purcell1948,Garbacz_1965_TCM,Harrington+Mautz1971} plays an important role in the design~\cite{lau2022characteristic} of antennas, such as electrically small antennas, MIMO systems, and arrays. A recent extension of the scattering-based formulation of characteristic mode decomposition~\cite{Garbacz_1965_TCM} has broadened its application scope to include arbitrary electromagnetic solvers~\cite{Gustafsson+etal2022a,Gustafsson+etal2022b,Capek+etal2023a}, enabling advanced applications with arbitrary material distributions. Despite its numerous advantages~\cite{Capek+etal2023a} and rapid evaluation capabilities~\cite{Lundgren+etal2023}, the scattering approach to characteristic modes was not readily extended to the substructure variant~\cite{Ethier+McNamara2012} frequently applied using impedance-based methods~\cite{Harrington+Mautz1971}.

Substructure characteristic mode decomposition involves dividing the scattering scenario into a controllable (or accessible, see~\cite{Ethier+McNamara2012,alroughani2013appraisal}) \TB{design} region and an \TB{uncontrollable} background. In this approach, any structural modification, such as antenna design or selective excitation, is confined to the controllable region. An example is a patch antenna situated over a ground plane, where the ground plane has a significant impact on most characteristic modes, yet the designer can effectively influence only the \TB{patch ~\cite{Angiulli+etal2000,Ethier+McNamara2012, 2023_Zhao_TAP}}. The substructure characteristic mode decomposition focuses on modes most closely associated with the controllable region by using altered forms of operators describing the scattering problem. This makes substructure modes an attractive approach for studying the behavior of radiating devices affected by nearby objects like vehicles, electronic platforms, or biological tissues~\cite{Luomaniemi_CMA2021}.

Following the state-of-the-art procedures, substructure characteristic modes can be interpreted as a classical formulation of characteristic mode decomposition where the underlying Green's dyadic includes the surroundings (background) of the studied region~\cite{Parhami+etal1977,cwik1989constructing,dai2016characteristic, alakhras2020sub}. For a few special cases, the Green's dyadic can be computed analytically (e.g., infinite ground planes), however most practical problems require numerical solutions. One implementation of this approach using the method of moments (MoM)~\cite{Harrington1968} formulations is based on a Schur complement~\cite{Ethier+McNamara2012, 2014_Alroughani_ICEAA}, which constructs a compressed impedance matrix for a scatterer in the presence of background objects. Several variations of this technique have been developed to account for a variety of specific background problems, all following the Schur complement method applied to various integral equation formulations~\cite{wu2019general,wu2019characteristic,boyuan2021sie,2021_Huang_TAP,wu2022characteristic,guo2022novel,2023_Huang_TAP,guo2024substructure,2024_Guo_TAP}.

This paper proposes an alternate approach to the computation of substructure modes for lossless problems without the explicit requirement of integral equations by introducing a scattering-based variant, formulated through a generalized eigenvalue problem of two scattering matrices~\cite[\S~7.8.1]{Kristensson2016}, \cite[\S~4.3]{Pozar2005}. The first matrix accounts for the complete scattering problem, while the second represents the background. \TB{Unlike previous methods, this approach is general and opens up the analysis of previously inaccessible scenarios such as bi-anisotropic medium and ports (see Sec.~\ref{sec:discu}).}
The employed scattering matrices can be easily substituted with transition matrices (T-matrices)~\cite{Mishchenko+etal1996} or by scattering dyadics~\cite[\S~4.3]{Kristensson2016}, which can be constructed by an arbitrary numerical technique~\cite{Gustafsson+etal2022a,Gustafsson+etal2022b,Capek+etal2023a}. Furthermore, the proposed algorithm can be accelerated through the use of an adapted form of a previously-reported iterative algorithm~\cite{Lundgren+etal2023}, which mitigates the computational burden associated with increasing electrical size and model complexity in substructure problems. 

The scattering-based and MoM-based substructure formulations are shown to be identical \TB{ in the sense that they yield the same modes in cases where both methods are applicable}. Nevertheless, there are cases where the MoM-based variant is not easily applicable or not advantageous to use. This makes the proposed formulation applicable to various problems typically outside the scope of characteristic mode analysis, including microwave circuits and optical circuits spanning large ranges of electrical sizes with complex material distributions. In summary, the proposed method inherits all of the properties of the scattering formulation of characteristic modes \cite{Garbacz_1965_TCM,Gustafsson+etal2022a} including the most impactful generalization that if the scattering response of an object and its background can be computed (by any means), then substructure characteristic modes can be computed, regardless of the background or object's complexity and whether or not integral equations were used.

The paper is organized as follows. Section~\ref{sec:substr} briefly reviews the basic concept of 
substructure characteristic modes. 
Section~\ref{sec:scatForm} introduces the scattering-based formulation and its equivalence to MoM-based substructure characteristic mode decomposition is shown in Section~\ref{S:MoMsubstructure}. Several examples are presented, with emphasis on \TB{those} which can be computed both by previous methods and the proposed scattering-based technique.  These include \TB{a }simple object made of perfect electric conductor \TB{ (PEC)}, a hybrid of the method of moments and T-matrix method, and infinite ground plane. Computationally challenging examples are shown in Section~\ref{sec:extSolver}. The outcomes are discussed in Section~\ref{sec:discu}, and the paper is concluded in Section~\ref{sec:conclu}. 

\section{Substructure Characteristic Modes}
\label{sec:substr}

\begin{figure}[]
    \centering
    \includegraphics[width=\columnwidth]{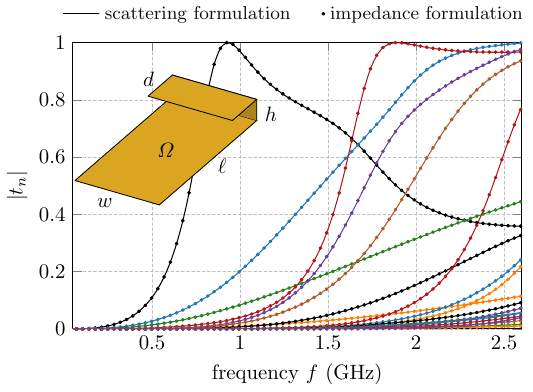}
    \caption{Modal significances $|t_n|$ computed for a PEC geometry adapted from~\cite{Ethier+McNamara2012} using the scattering-based formulation~\eqref{eq:Seig} with $\M{S}_\bSub$ being unit matrix, \ie{}, $\M{S}_\bSub = \M{1}$, (solid lines), and the impedance-based formulation (dots)~\cite{Harrington+Mautz1971}.  The dimensions read~\mbox{$\ell = 120\,\T{mm}$}, \mbox{$w = 60\,\T{mm}$}, \mbox{$h = 15\,\T{mm}$}, and \mbox{$d = 30\,\T{mm}$}.}
    \label{fig:ethier-full}
\end{figure}

\begin{figure}
    \centering
    \includegraphics[width=\columnwidth]{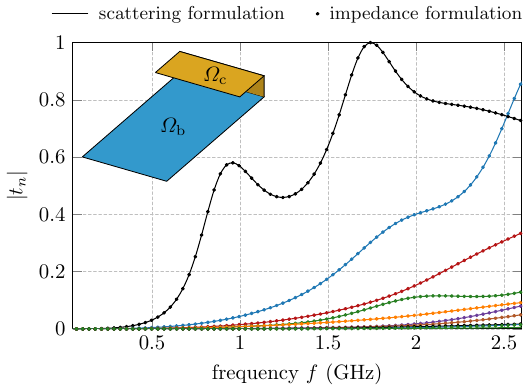}
    \caption{Modal significances $|t_n|$ computed for the substructure $\regc$ in the presence of the background scatterer $\regu$. Both regions are PEC with dimensions identical to the structure in Fig.~\ref{fig:ethier-full}. The impedance-based method relies on formulation~\cite{Ethier+McNamara2012}, while the scattering-based formulation relies on~\eqref{eq:Seig}.}
    \label{fig:ethier-sub}
\end{figure}

Insights into essential features of substructure characteristic modes can be obtained by comparing Fig.~\ref{fig:ethier-full} and  Fig.~\ref{fig:ethier-sub}, where we adapt a PIFA-like example~\cite{Ethier+McNamara2012} made of \TB{PEC}\footnote{We study variations of this example throughout the remainder of the manuscript using its original dimensions and frequency analysis range from~\cite{Ethier+McNamara2012}, though the absence of losses means that all problems can be scaled in physical size to cover arbitrary frequency ranges without alteration to the theoretical technique or numerical results.}, illustrating the characteristic modal significance of the entire region~$\reg=\regc\cup\regu$ and substructure modal significance corresponding to the controllable region $\regc$ in the presence of the background region~$\regu$.  

In Fig.~\ref{fig:ethier-full}, the entire device is studied, and the resulting modal spectrum includes modes which can be induced and superimposed by characteristic excitations applied over the entire structure.  This results in a dense set of modes with high modal significance.  However, when excitations are confined to the design region $\regc$, treating the region $\regu$ as a fixed background, the resulting modal spectrum, shown in Fig.~\ref{fig:ethier-sub}, exhibits notable differences. Here, only modes that can be induced or superimposed using excitations on the design region~$\regc$ are present, resulting in a much sparser eigenspectrum and revealing a resonance peak around~$1.8 \, \T{GHz}$ which is supported by currents on the PIFA.  It is important to note that, in this case, modal current distributions exist over both regions $\regc$ and $\regu$. Nevertheless, the absence of independent control of excitations over the region $\regu$ effectively removes currents over that region from the set of degrees of freedom defining each mode. 

Previously, calculations like these were limited by their reliance on MoM~\cite{Ethier+McNamara2012,capek2022computational}. This paper demonstrates that substructure characteristic modes can be evaluated using the scattering matrix formalism, significantly expanding their scope and computational capabilities.

\section{Scattering Formulation of Substructure Characteristic Modes}
\label{sec:scatForm}

Assume a controllable antenna region $\regc$ and the surrounding (or background) region $\regu$, as illustrated in Fig.~\ref{fig:subCMscatt}. Let~$\M{S}_\bSub$ denote that scattering matrix~\cite{Pozar2005,Kristensson2016} for the background region~$\regu$, which connects incoming waves represented by coefficients collected in a vector~$\M{a}$ into outgoing waves represented by a vector~$\M{f}$. Similarly, let $\M{S}$ denote the scattering matrix for the composite object $\regc\cup\regu$. These scattering matrices are general, and the basis functions used to expand the incident and reflected waves depend on the specific problem. For a free-space scattering problem it is convenient, though not necessary, to use spherical vector waves as basis functions~\cite{Kristensson2016}. In this case, the column vectors $\M{f}$ and $\M{a}$ contain the spherical wave expansion coefficients, similar to a multipole expansion.

The core hypothesis of this paper is that substructure characteristic modes are determined from the generalized eigenvalue problem~\cite{Gustafsson+etal2023}
\begin{equation}
    \M{S}\M{a}_n=s_n\M{S}_\bSub\M{a}_n,
		\label{eq:Seig}
\end{equation}
where~$\M{a}_n$ are characteristic excitations and~$s_n$ characteristic scattering eigenvalues. Characteristic scattered fields are represented by vectors $\M{f}_n=\M{S}_\bSub\M{a}_n$ which collect expansion coefficients of the scattered field in the same basis.

The dimension of the square matrices $\M{S}$ and $\M{S}_\T{b}$ are the same, with conservative estimates available for the number of entries based on the electrical size of the problems being studied\footnote{If the background structure is much smaller than that of the controllable region, then a smaller dimension can be used to accelerate computation of the background scattering matrix $\M{S}_\T{b}$ before augmenting it with an identity matrix to match the dimension of matrix~$\M{S}$ which represents the complete structure.}, such as the electrical radius for expansions in spherical waves~\cite{Song+Chew2001b,Tayli+etal2018}. This formulation generalizes the one defined by~\cite{Garbacz+Turpin1971} and developed in~\cite{Gustafsson+etal2022a} for objects in free space, which implicitly considers~the background scattering matrix $\M{S}_\bSub$ to be an identity matrix.  

The formulation above also covers the special case studied for periodic structures in~\cite{Schab+etal2023}, where the background scattering matrix $\M{S}_\bSub$ is related to an implicit connection between plane waves (Floquet modes) on either side of a scattering surface, analogous to the S-parameters of a transmission line network. The background problem for periodic systems can also be modified to include a ground plane or supporting material layers~\cite{Schab+etal2023} and the use of periodic boundary conditions leads to much more restrictive guidelines on the appropriate dimension of the scattering matrices based on the number of propagating Floquet modes~\cite{schab2021sparsity}.

The eigenvalues $s_n$ are related to modal significances $|t_n|$ and characteristic eigenvalues $\lambda_n$ as~\cite{Gustafsson+etal2022a}
\begin{equation}
	t_n = \frac{s_n-1}{2}
	\quad\text{and }
	\lambda_n = \T{j}(1+t_n^{-1}) = \ju\frac{s_n+1}{s_n-1}.
\label{eq:tn}
\end{equation}
Similarly to characteristic modes of isolated objects~\cite{Gustafsson+etal2022a}, for lossless objects, characteristic modes exhibit the following orthogonality relations
\begin{equation}
    \M{a}_m^{\herm}\M{a}_n=\delta_{mn},\quad  
    \M{f}_m^{\herm}\M{f}_n=\delta_{mn},
\end{equation}
which, for example, translates to the orthogonality of characteristic far fields and where~$^\T{H}$ denotes Hermitian transpose and~$\delta_{mn}$ is the Kronecker delta. Equivalence between \TB{the modes from the }scattering and impedance formulations is supported by data shown in Figs.~\ref{fig:ethier-full} and~\ref{fig:ethier-sub} comparing results from modal significances produced by \eqref{eq:Seig} and an impedance-based formulation of substructure modes~\cite{Ethier+McNamara2012}. A mathematical derivation of this equivalence is outlined in Section~\ref{S:MoMsubstructure} and Appendix~\ref{S:SS0MoMeq}.


Throughout the remainder of the paper, we use the quantities $\M{a}_n$ and $s_n$ as representations of characteristic modes due to their \TB{ direct relation to} the characteristic currents~$\M{I}_n$ typically used in impedance-based formulations~\cite{Harrington+Mautz1971,Gustafsson+etal2022a}. These characteristic currents are typically referred to as "excitation independent", but they can always be driven by an appropriate excitation $\M{V}_n = \M{Z}\M{I}_n$ (obtained using the impedance matrix~$\M{Z}$) or, equivalently, the superposition of spherical waves described by the vector $\M{a}_n$. Details on recovering the characteristic currents are given in Section~\ref{sec:IVa}. Other characteristic quantities, such as fields, are computed by \TB{exciting the system with an excitation described by $\M{a}_n$ and measuring the quantity of interest, using whatever solver was used to construct the scattering matrices $\M{S}$ and $\M{S}_\T{b}$.  The explicit mapping between incident fields and modal currents within MoM-based solvers is discussed in Section~\ref{S:MoMsubstructure}.} 

\tikzstyle{myHatch} = [pattern=north west lines, pattern color=gray, line width=0.1pt, draw opacity=1]
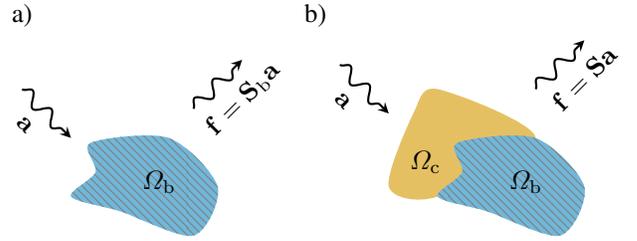
\begin{figure}%
{\centering
\begin{tikzpicture}[scale=0.65,>= stealth]
\begin{scope}
\node at (-2, 2.5) {a)};
\begin{scope}[xshift=-1cm]
\fill[breg!70,very thick] plot [smooth cycle,scale=1] coordinates {(0, -1.2) (0.5, -0.7) (0.4, -0.1) (2,0) (3,-1) (2.5, -2) (1,-1.5)};
\fill[myHatch] plot [smooth cycle,scale=1] coordinates {(0, -1.2) (0.5, -0.7) (0.4, -0.1) (2,0) (3,-1) (2.5, -2) (1,-1.5)};			
\node at (1.8,-0.9) {$\regu$};
 \end{scope}
 
\draw[->,thick,decorate,decoration=snake] (-2,1) -- node[sloped,below, xshift=-0.1cm, yshift=-0.1cm] {$\M{a}$} (-1,0);
\draw[->,thick,decorate,decoration=snake] (1.5,0.6) -- node[sloped,below, xshift=0.1cm, yshift=-0.1cm] {$\M{f}=\M{S}_\bSub\M{a}$} +(1,1);	
\end{scope}

\begin{scope}[xshift=6.5cm]
\node at (-2.5, 2.5) {b)};
\fill[reg,very thick, opacity=\opC] plot [smooth cycle,scale=1] coordinates {(-1,-1) (0,-1.2) (2,0) (0,1) (-0.5,0.5)};
\node[] at (-0.25, -0.5) {$\regc$};
\fill[breg!70,very thick] plot [smooth cycle,scale=1] coordinates {(0, -1.2) (0.5, -0.7) (0.4, -0.1) (2,0) (3,-1) (2.5, -2) (1,-1.5)};
\fill[myHatch] plot [smooth cycle,scale=1] coordinates {(0, -1.2) (0.5, -0.7) (0.4, -0.1) (2,0) (3,-1) (2.5, -2) (1,-1.5)};			
\node at (1.8,-0.9) {$\regu$};

\draw[->,thick,decorate,decoration=snake] (-2,1.5) -- node[sloped,below, xshift=-0.1cm, yshift=-0.1cm] {$\M{a}$} (-1,0.5);	
\draw[->,thick,decorate,decoration=snake] (2,1) -- node[sloped, below, xshift=0.1cm, yshift=-0.1cm] {$\M{f}=\M{S}\M{a}$} +(1,1);		
\end{scope}
\end{tikzpicture}\par}
\caption{Sketch of the physical interpretation of scattering matrices~$\M{S}_\bSub$ and~$\M{S}$.}%
\label{fig:subCMscatt}%
\end{figure}

\subsection{Variations}
For lossless objects, the eigenvalue problem in \eqref{eq:Seig} can be rearranged into several alternative forms. 
 Scattering matrices are unitary for lossless objects~\cite{Pozar2005}, $\M{S}^{\herm}\M{S}=\M{1}$, leading to
\begin{equation}
    \M{S}_\bSub^{\T{H}}\M{S}\M{a}_n=s_n\M{a}_n    
    \quad\text{and}\quad
    \M{S}\M{S}^{\T{H}}_\bSub\M{f}_n=s_n\M{f}_n,
		\label{eq:Seigalt}
\end{equation}
where $\M{f}_n=\M{S}_\bSub\M{a}_n=s_n^{\ast}\M{S}\M{a}_n$ represents the scattered field, $\M{1}$ is the identity matrix, and $^\ast$~denotes complex conjugate. The two versions in~\eqref{eq:Seigalt} are equivalent and differ only in expressing the eigenvalue problem in incoming (excitation), $\M{a}_n$, or outgoing (scattered or radiated), $\M{f}_n$, waves. We note that the incident field $\M{a}_n$ differ from the scattered field $\M{f}_n$ except for the free-space case with $\M{S}_\bSub=\M{1}$~\cite{Gustafsson+etal2022a}.

Formulation~\eqref{eq:Seigalt} can be implemented using transition matrices\footnote{Alternatively, scattering dyadics~\cite{Kristensson2016} can be used in place of the transition matrices $\M{T}$ and $\M{T}_\bSub$ with no further changes to the formulation presented in the remainder of this section, save for the understanding that the transition matrix maps regular spherical waves to outgoing spherical waves, while the scattering dyadic maps incoming plane waves to the scattered far field. 
See \cite{Capek+etal2023a} for details on these two operators and their use in characteristic modes.}~\cite{Kristensson2016}
\begin{equation}
\M{S}=2\M{T}+\M{1}\quad\text{and}\quad \M{S}_\bSub=2\M{T}_\bSub+\M{1},
\label{eq:S2T}
\end{equation}
which allows for the analysis of substructure characteristic modes solely in terms of transition matrices and characteristic numbers~$t_n$. These formulations read
\begin{equation}
   (2\M{T}_\bSub^{\T{H}}\M{T}+\M{T}_\bSub^{\T{H}}+\M{T})\M{a}_n
    =t_n\M{a}_n    
	\label{eq:TeigA}
\end{equation}
and 
\begin{equation}
   (2\M{T}\M{T}_\bSub^{\T{H}}+\M{T}_\bSub^{\T{H}}+\M{T})\M{f}_n
    =t_n\M{f}_n,
    \label{eq:Teigf}
\end{equation}
which involve products of transition matrices, unlike the characteristic mode formulation of isolated objects~\cite{Gustafsson+etal2022a}. In contrast to~\eqref{eq:Seig}, relations~\eqref{eq:Seigalt},~\eqref{eq:TeigA}, and~\eqref{eq:Teigf} enable iterative matrix-free evaluation~\cite{Lundgren+etal2023}, an important factor when employing generic electromagnetic solvers for evaluating substructure characteristic modes for large problems. Details of the iterative solution are found in~Appendix~\ref{S:ItCMsub}. Moreover, relation~\eqref{eq:TeigA} allows for interpreting the substructure scattered power as a power of the difference between the scattered field of the composite object $\M{T}\M{a}_n$ and the scattered field of the background $\M{T}_\bSub\M{a}_n$, \ie
\begin{equation}
	\frac{1}{2}|(\M{T}-\M{T}_\bSub)\M{a}_n|^2
	=\frac{-\Re\{t_n\}}{2}|\M{a}_n|^2
	=\frac{1}{2}|t_n|^2|\M{a}_n|^2 
\label{eq:CMmaxscatt}
\end{equation}
with the interpretation of no substructure scattering for $t_n=0$ (or $s_n=1$) and maximum scattering for $t_n=s_n=-1$ similar to the free-space case~\cite{Gustafsson+etal2022a}.  Here, we have utilized the property~\cite{Gustafsson+etal2022a}
$\M{T}^\T{H}\M{T} = -\T{Re}\{\M{T}\}$.


\section{Equivalence Between MoM and Scattering-Based Substructure Characteristic modes}
\label{S:MoMsubstructure}

\TB{The proposed formula~\eqref{eq:Seig} yields the same modes as the impedance-based substructure formulation}~\cite{Ethier+McNamara2012}, wherein a structure is separated into two (possibly connected) regions~$\regu$ and~$\regc$, and excitation is restricted to only the controllable region~$\regc$~\cite{Gustafsson+etal2023}.  A general scattering problem in the standard matrix representation of surface and / or volumetric electric field integral equation (EFIE)-formulations~\cite{Harrington1968} reads
\begin{equation}
    \Vvec = \Zmat\Ivec,
\end{equation}
where $\Vvec$ is an excitation vector, $\Zmat$ is the impedance matrix associated with the scatterer~$\reg$, and $\Ivec$ is an induced current density, all represented in a particular basis.  Bifurcating the system into subsets of the basis functions associated with the two regions~$\regu$ and~$\regc$ and enforcing zero excitation on the background region leads to~\cite{Parhami+etal1977}
\begin{equation}
\Zmat \Ivec = 
	\begin{bmatrix}
		\M{Z}_{\bSub\bSub} & \M{Z}_{\bSub\cSub} \\
		\M{Z}_{\cSub\bSub} & \M{Z}_{\cSub\cSub} 
	\end{bmatrix}
\begin{bmatrix}
		\M{I}_\bSub \\
		\M{I}_\cSub 
	\end{bmatrix}	
	=
\begin{bmatrix}
		\M{0}\\
		\M{V}_\cSub 
	\end{bmatrix}.
\label{eq:SubMoM}
\end{equation}
By reducing the MoM system $\M{Z}=\M{R}+\ju\M{X}$ to its Schur complement, the substructure characteristic modes are defined as~\cite{Ethier+McNamara2012}
\begin{equation}
\widetilde{\M{X}}\M{I}_{\cSub n} = \lambda_n\widetilde{\M{R}}\M{I}_{\cSub n}
	\quad\text{or} \quad
	\widetilde{\M{Z}}\M{I}_{\cSub n} = (1+\ju\lambda_n)\widetilde{\M{R}}\M{I}_{\cSub n},
\label{eq:CMsubMoM}
\end{equation}
where $\widetilde{\M{Z}} = \widetilde{\M{R}} + \J \widetilde{\M{X}} = \M{Z}_{\cSub\cSub}  - \M{Z}_{\cSub\bSub}\M{Z}_{\bSub\bSub}^{-1}\M{Z}_{\bSub\cSub}$ is an impedance matrix related to the problem-specific numerical Green's function~\cite{Parhami+etal1977}. This numerical Green's function captures the field-generating behavior of currents on the controllable region~$\regc$ in the presence of the background region~$\regu$.

The proof of equality of~\eqref{eq:CMsubMoM} and~\eqref{eq:Seig} starts with factorization of the radiation matrix~\cite{Tayli+etal2018}
$\M{R} = \M{U}_1^\trans \M{U}_1$, where the matrix~$ \M{U}_1$ projects MoM basis functions onto spherical waves. Next step is partitioning~$\M{U}_1 = \begin{bmatrix}
    \M{U}_{1\bSub} & \M{U}_{1\cSub}
\end{bmatrix}$
and factorization
\begin{equation}
\widetilde{\M{R}} = \widetilde{\M{U}}_1^\herm \widetilde{\M{U}}_1 
\label{eq:Rtil}
\end{equation}
for the substructure case, where $\widetilde{\M{U}}_1 = \M{U}_{1\cSub} - \M{U}_{1 \bSub} \M{Z}_{\bSub\bSub}^{-1}\M{Z}_{\bSub\cSub}$. A substructure eigenvalue problem for T-matrix
\begin{equation}
	\wtM{T}\M{f}_n = t_n \M{f}_n
\label{eq:CMeigUZU}	
\end{equation}
with~$t_n = \left(1 + \T{j} \lambda_n \right)^{-1}$ can be formulated by substituting~\eqref{eq:Rtil} into~\eqref{eq:CMsubMoM}, left multiplication with~$\widetilde{\M{U}}_1 \wtM{Z}^{-1}$ and by identifying
\begin{equation}
    \wtM{T} = -\wtM{U}_1\wtM{Z}^{-1}\wtM{U}_1^{\herm} \quad\T{and}\quad \M{f}_n = - \widetilde{\M{U}}_1 \M{I}_{\cSub n}.
    \label{eq:Ttil}
\end{equation}
Appendix~\ref{S:SS0MoMeq} then shows that matrix~$\wtM{T}$ equals to the matrix used in~\eqref{eq:Teigf}. 

The characteristic current can be obtained from the characteristic excitation as
\begin{equation}
     \M{I}_n = \begin{bmatrix}
		\M{I}_{\bSub n} \\
		\M{I}_{\cSub n} 
	\end{bmatrix}	 = \M{Z}^{-1}\M{U}_1^{\herm} \M{a}_n - \begin{bmatrix}
		 \M{Z}_{\bSub\bSub}^{-1} \M{U}_{1 \bSub}^{\herm}   \\
		\M{0} 
	\end{bmatrix} \M{a}_n,	
 \label{eq:eigenCurrent1}
\end{equation}
\ie{}, by subtracting the current induced by excitation~$\M{a}_n$ on the background object from the current induced
on the composite object. Employing the abbreviations introduced in this section, relation~\eqref{eq:eigenCurrent1} is identical to 
\begin{equation}
 \M{I}_{\cSub n} = \dfrac{1}{t_n} \wtM{Z}^{-1}\wtM{U}_1^{\herm}\M{f}_n. 
\end{equation}

\subsection{Example --- Equivalence and Substructure Interpretation}
\label{sec:IVa}
To demonstrate the numerical equivalence between the \TB{modes from the }impedance and scattering formulations in \eqref{eq:CMsubMoM} and \eqref{eq:Seig}, we employ a PEC structure reported in \cite{Ethier+McNamara2012}, which consists of a PIFA-like region connected to a finite ground plane, as depicted in the inset of Fig.~\ref{fig:ethier-full}.  In Fig.~\ref{fig:ethier-full}, the characteristic modes of the full structure are computed without any separation into controllable and background regions, \ie{}, the entire region is controllable. Impedance and scattering operators are calculated using MoM, and the modal significances produced by the impedance and scattering formulations are numerically identical.  Similarly, Fig.~\ref{fig:ethier-sub} displays modal significances when only part~$\regc$ of the structure is controllable (following \cite{Ethier+McNamara2012}), and again the impedance~\eqref{eq:CMsubMoM} and scattering formulations~\eqref{eq:Seig} agree within numerical precision.

\begin{figure}
    \centering
    \includegraphics[width=\columnwidth]{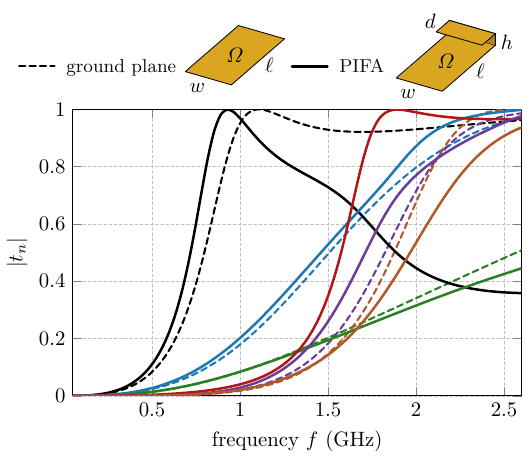}
    \caption{Comparison between modal significances $|t_n|$ for the PIFA structure in Fig.~\ref{fig:ethier-full} (solid) and a ground plane (dashed). No background is considered,  \ie{}, $\M{S}_\bSub = \M{1}$. The colors are paired according to the similarity in radiation diagrams. When evaluating substructure characteristic modes in Fig.~\ref{fig:ethier-sub}, the ground plane (dashed lines) is taken as a background. All dimensions are the same as in Fig.~\ref{fig:ethier-full}.}
    \label{fig:ethier-bg}
\end{figure}  

The transformation of full-structure modes, shown in Fig.~\ref{fig:ethier-bg}, into substructure modes in Fig.~\ref{fig:ethier-sub} can be intuitively explained using the excitation and radiated fields described in~\eqref{eq:CMmaxscatt}.

To illustrate this, consider the resonances around $1\unit{GHz}$ for both full structures in Fig.~\ref{fig:ethier-bg}. These resonances correspond to the fundamental mode of each structure where the far-field closely resembles that of an electric dipole, with characteristic currents flowing along the longest dimension. The resonant frequency of the full PIFA structure is slightly lower than that of the ground plane alone, mainly due to the increased effective length of the overall configuration.

Assuming similar characteristic excitation $\M{a}_n$ for both modes near resonance, \eqref{eq:CMmaxscatt} suggests that substructure characteristic modes result from the difference in scattering -- both amplitude and phase -- between the full PIFA structure and the ground plane ``background''. This difference is reflected in the substructure eigenvalues in Fig.~\ref{fig:ethier-sub}, which is given from the differences of the eigenvalues in Fig.~\ref{fig:ethier-bg}. Consequently, no resonance is observed near $1 \unit{GHz}$ in Fig.~\ref{fig:ethier-sub}; however, the dominant substructure mode exhibits a peak with a modal significance of $|t_n|\approx 0.6$. This behavior is attributed to the frequency shift between the black lines in Fig.~\ref{fig:ethier-bg} and the resulting imperfect cancellation of eigenvalues.

The second characteristic mode, shown by the blue traces in Fig.~\ref{fig:ethier-bg}, similarly has a far field characterized by that of an electric dipole, with characteristic currents flowing along the width dimension. The characteristic traces are nearly identical for both structures, resulting in a significant reduction for the substructure modal significance, as indicated by~\eqref{eq:CMmaxscatt}. Finally, the resonance of the full structure in Fig.~\ref{fig:ethier-bg} around~$1.6\unit{GHz}$ is not supported by the ground plane (see the full red line in Fig.~\ref{fig:ethier-bg}) and is therefore replicated in the substructure data in Fig.~\ref{fig:ethier-sub}.  

These simple interpretations for the substructure characteristic modes rely on the assumption that the excitations $\M{a}_n$ are approximately the same for the entire structure and the background structure. This approximation holds well over large bandwidths for electrically small structures but only for narrow bandwidths for electrically large structures.


\subsection{Computational complexity}
\label{sec:comp-comp}
In contrast to the MoM matrix formulation~\eqref{eq:CMsubMoM}, the scattering-based substructure formulation~\eqref{eq:Seig} is solver-agnostic. The scattering formulation can thus take advantage of the rapid advances in computational electromagnetics (CEM) to solve scattering problems and, hence, effectively evaluate characteristic modes. A theoretical comparison of the computational complexity between the classical MoM formulation and the scattering-based formulation is presented for two idealized cases in Table~\ref{tab:CompComp}.

\begin{table}
\caption{Comparison between memory usage and number of operations for the MoM formulation~\eqref{eq:SubMoM} together with~\eqref{eq:CMsubMoM} and iterative matrix-free MoM formulations~\cite{Chew+etal2001,Polimeridis+etal2014} for the scattering formulation~\eqref{eq:Seig}.}
\begin{tabular}{cccc} 
	& & MoM & Scattering \\ \toprule
	\multirow{2}{*}{\includegraphics[width=2cm]{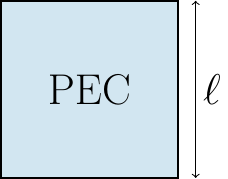}} & mem. & $(k\ell)^4$  & $(k\ell)^2$ \rule{0pt}{0.7cm}\\
	& ops. & $(k\ell)^6$ & $N_\T{cm}N_\T{it}(k\ell)^2\log (k\ell)$ \rule{0pt}{0.85cm}\\[0.5cm] \midrule
	\multirow{2}{*}{\includegraphics[width=2.2cm]{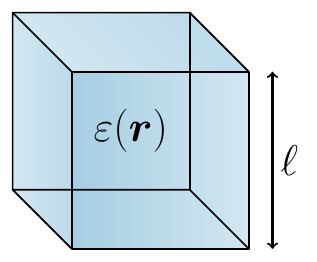}} & mem. & $(k\ell)^6$  & $(k\ell)^3$ \rule{0pt}{0.55cm}\\
	& ops. & $(k\ell)^9$ & $N_\T{cm}N_\T{it}(k\ell)^3\log (k\ell)$  \rule{0pt}{0.75cm} \\[0.7cm] \bottomrule
\end{tabular}
\label{tab:CompComp}
\end{table}

The background object in the first case consists of a PEC plate with side lengths $\ell$, while the second case features a dielectric cube with side lengths $\ell$ and a spatially dependent relative permittivity~\TB{$\varepsilon_\T{r}(\V{r})$}. For simplicity, the controllable region is assumed to be a square PEC plate with side lengths $\ell_\T{c}\ll\ell$. Assuming a discretization using 10 points per wavelength results in approximately $N_\T{b} \approx 10 \ell_\T{b}/\lambda\approx 1.6 k\ell$ free-space discretization points per side.
The memory usage of the MoM for the PEC plate is $N_\T{b}^4\sim(k\ell)^4$ matrix elements for $\M{Z}_{\T{b}\T{b}}$. The computational complexity for evaluation of the Schur complement~\eqref{eq:CMsubMoM} is proportional to $(k\ell)^6$. For the dielectric cube, the memory and complexity are proportional to $(k\ell)^6$ and $(k\ell)^9$, respectively, assuming the use of a volumetric electric field integral equation.

In contrast, the scattering problems underlying~\eqref{eq:Seig} can be solved using arbitrary solvers. Taking advantage of matrix-free MoM formulations, such as the Multi-Level Fast Multipole Method (MLFMM), reduces the memory requirement to $(k\ell)^2$~\cite{Chew+etal2001} for the PEC case. The characteristic modes are solved iteratively as outlined in Appendix~\ref{S:ItCMsub}, giving a computational complexity proportional to the number of evaluated characteristic modes $N_\mathrm{cm}$, MLFMM iterations $N_\T{it}$, and $(k\ell)^2\log(k\ell)$. The corresponding heterogeneous dielectric cube can, for instance, be solved using FFT-based volumetric MoM~\cite{Polimeridis+etal2014}, with a memory requirement of $(k\ell)^3$ and similar computational complexity.

Both cases in Table~\ref{tab:CompComp} demonstrate the unfavorable memory and computational complexity of MoM systems~\eqref{eq:SubMoM} for electrically large structures. The memory scaling of $(k\ell)^6$ for volumetric objects makes the MoM system~\eqref{eq:SubMoM} impractical for objects larger than a few wavelengths. In contrast, the full range of modern CEM techniques (including MoM, FEM, and FDTD) is directly applicable to the scattering-based formulation~\eqref{eq:Seig}. This example illustrates some advantages of~\eqref{eq:Seig} over the classical approach~\eqref{eq:CMsubMoM}. However, it is important to note that we do not claim that the scattering-based approach is universally superior. There are numerous situations where the classical method is computationally efficient and may be preferable, such as for electrically small controllable and background objects represented by a low number of basis functions (\eg, the example shown in Fig.~\ref{fig:ethier-sub}). The scaling in Table~\ref{tab:CompComp} is based on the use of a Schur complement~\eqref{eq:CMsubMoM} for the MoM matrices. \TB{However, recent advancements can potentially improve the computational efficiency of the impedance-based substructure characteristic modes~\cite{Wu+Wu2024}}, similar to how the MLFMM is applied to full characteristic modes in~\cite{Dai+etal2016}.

\subsection{MoM and T-matrix Hybrid}
\label{sec:TMoMhybrid}

Hybridization of MoM and T-matrix techniques efficiently models complex inhomogeneous structures, so long as the regions described by each method can be separated by a plane~\cite{Losenicky+etal2022}.  Transition matrices can be computed for arbitrary background problems, but the computational acceleration over full MoM implementations is most pronounced when analytic forms of transition matrices can be employed, \eg{}, Mie series results for layered spherical structures or infinite ground planes.  In this section, we discuss how this form of hybridization also allows for the efficient computation of substructure modes when part or all of the background region is represented by a transition matrix.

\begin{figure}
    \centering
    \includegraphics[width=\columnwidth]{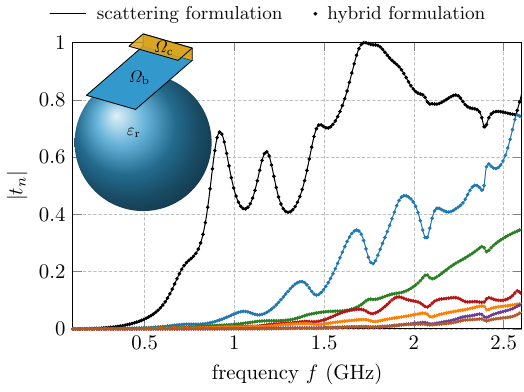}
    \caption{Modal significances~$|t_n|$ computed for the substructure~$\regc$ in the presence of the background which contains scatterer~$\regu$ and a dielectric sphere filled with relative permittivity~$\varepsilon_\T{r} = 4$. The diameter of the sphere equals~180~mm and the distance between the sphere and the metallic structure is 30~mm. Other dimensions are identical to Fig.~\ref{fig:ethier-full}. The impedance-based method relies on~\eqref{eq:CMsubMoM} and~\eqref{eq:Zhybrid}, while the scattering-based formulation relies on~\eqref{eq:Seig}.}
    \label{fig:hybridCMs}
\end{figure}

Consider a case from Fig.~\ref{fig:ethier-sub} to which a dielectric object is introduced as in~Fig.~\ref{fig:hybridCMs}. The scattering properties of a dielectric object can be efficiently described by matrix~$\M{T}_\T{\bSub 1}$, while the scattering properties of the metallic structure are described by the impedance matrix~$\M{Z}$ composed according to~\eqref{eq:SubMoM}. 

The reaction of the entire setup to an external excitation~$\M{a}^{\T{i}}$ can be written as~\cite{Losenicky+etal2022} 
\begin{equation}
	(\M{Z} + \M{U}^{\trans}_4\M{T}_\T{\bSub 1}\M{U}_4)\M{I}
	=(\M{U}_1^{\trans}+\M{U}_4^{\trans}\M{T}_\T{\bSub 1})\M{a}^{\T{i}}.
\label{eq:Zhybrid}
\end{equation}
The matrix~$\M{Z} + \M{U}^{\trans}_4\M{T}_\T{\bSub 1}\M{U}_4$ is the impedance matrix of a metallic body in the presence of a dielectric object, with the second term being interpreted as a contribution of matrix~$\M{T}_\T{\bSub 1}$ to the background Green's function~\cite{dai2016characteristic}. Characteristic decomposition of this modified impedance matrix would lead to characteristic modes of the entire metallic body in the presence of a dielectric sphere~\cite{2022_Jelinek_APS}. On the other hand, if substructure modes solely excitable from region~$\regc$ are desired, then this matrix is further modified and decomposed according to~\eqref{eq:CMsubMoM}. The resulting spectrum of modal significances, where the entire region~$\regu$ (the sphere together with the bottom metallic plate) is considered a background, is shown in~Fig.~\ref{fig:hybridCMs}. The equivalence \TB{of the modes }of this formulation to~\eqref{eq:Seig} is demonstrated in the same figure. In the scattering formulation, the hybrid method~\cite{Losenicky+etal2022} was instead used to obtain scattered field from the excitations~$\M{a}$ and, therefore, to obtain the corresponding transition matrices~$\M{T}$ and $\M{T}_\T{b}$.

\subsection{Infinite Ground Plane}

\begin{figure}
    \centering
    \includegraphics[width=\columnwidth]{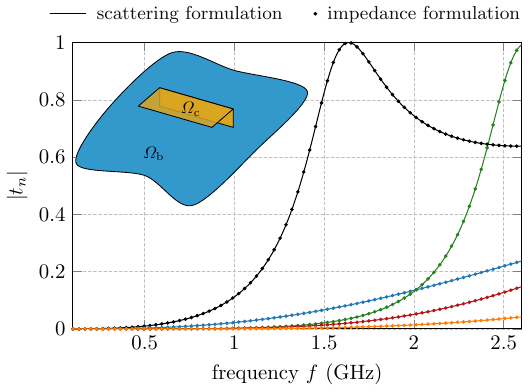}
    \caption{Modal significances $|t_n|$ computed for the substructure $\regc$ in the presence of the background $\regu$, which in this case is an infinite perfectly conducting ground plane. Dimensions are identical to Fig.~\ref{fig:ethier-full}.}
    \label{fig:ethierInfGND}
\end{figure}

An extreme version of the structure studied in Fig.~\ref{fig:ethier-sub} is a region~$\regc$ placed above an infinitely large perfectly conducting plane as depicted in Fig.~\ref{fig:ethierInfGND}.  When substructure characteristic modes are evaluated in this scenario, a common practice is to construct and decompose the matrix~$\widetilde{\M{Z}}_{\cSub\cSub}$ for the region~$\regc$ using the Green's function based on equivalent image currents~\cite{Angiulli+etal2000}. Another possibility is to decompose the impedance matrix belonging to region~$\regc$ and its mirror image and then employ point symmetries~\cite{Maseketal_ModalTrackingBasedOnGroupTheory} to filter out modes belonging to irreducible representation with even parity (those would belong to a perfectly magnetically conducting ground plane). 

Within the scattering formulation~\eqref{eq:Seig}, the problem is solved by adding an image of region~$\regc$ and an image of the incident field. This results in a total electric field with vanishing tangential components at the ground plane~$\regu$. The symmetry of the incident field eliminates half of the spherical waves. Specifically, only spherical waves exhibiting electric dipole moment normal to the ground plane and tangential magnetic dipole moment remain. For example, placing the ground plane~$\regu$ in the $xy$ plane and denoting~$l,m$ the degree and azimuthal numbers, respectively, only TE spherical waves with even $l+m$ and TM spherical waves with odd $l+m$ remain. \TB{The equivalence between modes produced by the impedance and scattering formulations is presented in Fig.~\ref{fig:ethierInfGND}.} 

The practicality of using substructure characteristic modes is demonstrated by comparing spectra from Figs.~\ref{fig:ethier-sub},~\ref{fig:hybridCMs} and~\ref{fig:ethierInfGND}. In all cases, the black resonance peak appearing between~$1.5 \, \T{GHz}$ and $2 \, \T{GHz}$ can be identified but would be lost if full characteristic decomposition was made, \cf{}, Fig.~\ref{fig:ethier-full}. This resonance belongs to a mode responsible for the PIFA operation, and substructure decompositions show how it is affected by a particular background.

\section{Utilization of Arbitrary Full-Wave Solver}
\label{sec:extSolver}

In case no in-house code is available to construct scattering matrices~$\M{S}$ and~$\M{S}_\bSub$, commercial simulators can be employed instead. This requires an interface between the simulator and a post-processor to assemble and decompose the scattering matrices.

To demonstrate the flexibility of scattering formulation~\eqref{eq:Seig}, the scattering dyadic matrices~\cite{Capek+etal2023a} are employed in this section to evaluate sub-structure characteristic modes on several complex examples using a commercial solver as the core computational engine. To mitigate the computational burden stemming from the fact that the full-wave evaluation is repetitively performed for all columns of the scattering dyadic matrices, an iterative procedure~\cite{Lundgren+etal2023} is employed and modified for the substructure case, see Appendix~\ref{S:ItCMsub}. Codes for these examples implemented in MATLAB and Altair FEKO~\cite{feko} are available at \cite{ScatSubstrFEKO}.

\subsection{Planar Inverted-F Antennas (PIFA)}

\begin{figure}
\centering
\includegraphics[width=\columnwidth]{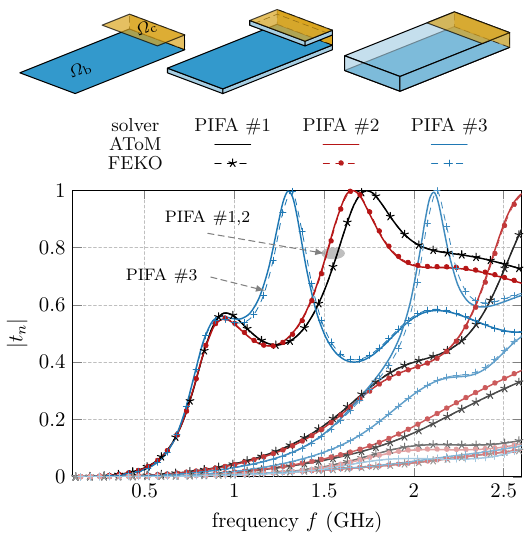}
\caption{The first five substructure characteristic modes for different planar inverted-F antennas. The first model is taken from Fig.~\ref{fig:ethier-sub} and serves as a verification. The second model is built on a substrate with a thickness of 1.575\,mm and relative permittivity~$\varepsilon_\T{r}=2.33$. The third model has a dielectric block of $\varepsilon_\T{r}=2.33$, filling the space between the ground plane and the antenna. All models are lossless. The controllable and background regions~$\regc$ and $\regu$ are depicted by the yellow and blue colors in the top pane, respectively. Full lines belong to the iterative algorithm that employs surface equivalence method-of-moments from Altair FEKO~\cite{feko}. The traces were adaptively refined to obtain smooth curves with relatively few frequency samples. The markers belong to solution via~\eqref{eq:Seig},  where scattering matrices were obtained by a solver~\cite{atom} combining surface and volumetric electric field integral equation. }
\label{fig:ethier-diel}
\end{figure}

A PIFA of the same dimensions as in Fig.~\ref{fig:ethier-full} is studied once more, now with the addition of two example configurations of dielectric substrates. Modal significance data for these examples, along with the previously studied PEC-only model, are shown in Fig.~\ref{fig:ethier-diel}. Five dominant modes are shown to emphasize the effect of the different dielectric configurations on detuning the fundamental resonance behaviors of the structure.

For comparison, markers are used in the same figure to present results of~\eqref{eq:Seig}, where the scattering matrices were obtained by a solver combining surface and volumetric electric field integral equation~\cite{atom}.

The reference case (PIFA~\#1) is made solely of PEC and represented by black traces. This case serves only as a verification of the scattering dyadic matrix procedure~\cite{Capek+etal2023a}. The resulting data are indistinguishable from the curves in Fig.~\ref{fig:ethier-sub}. The two cases involving dielectrics are also shown in Fig.~\ref{fig:ethier-diel} and include two configurations of a lossless dielectric material with relative permittivity $\varepsilon_\T{r}=2.33$. One case (PIFA~\#2) is built using thin dielectric layers, while the other case (PIFA \#3) includes a dielectric block spanning the entire space between the ground plane and the upper motif.

It can be seen that the eigen-traces of PIFA~\#1 and PIFA~\#2 are comparable, which is given by the fact that thin substrate reduces the effect of the dielectrics, \ie{}, the effective permittivity between the ground plane and the upper motif is close to unity, leading to only a slight downward shift in the resonant frequency of the dominant characteristic mode. The case with PIFA~\#3 behaves similarly, but the modal significance maxima are shifted considerably towards lower frequencies due to the reduced wavelength within the dielectric. The frequency dependence of the trace associated with the dominant mode also differs significantly past the resonance. 

By studying the detuning effects of different dielectric configurations, this example demonstrates just one potential analysis method afforded by the scattering formulation without restriction to a particular integral representation.

\begin{figure}
\centering
\includegraphics[width=\columnwidth]{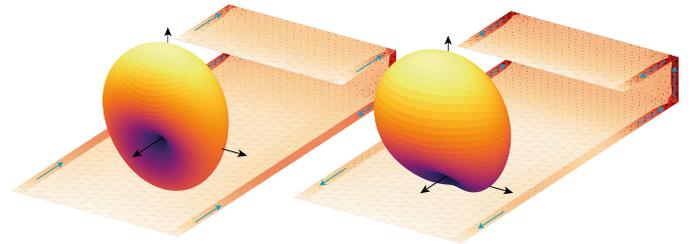}
\caption{Characteristic current density and characteristic directivity pattern of the dominant characteristic mode of PIFA~\#3. The left panel corresponds to the maximum at frequency~$f \approx 0.95 \, \T{GHz}$ and shows a maximum directivity~$D \approx 2.1$. The right panel corresponds to the maximum at frequency~$f \approx 1.3 \, \T{GHz}$ and shows a maximum directivity~$D \approx 3.1$.}
\label{fig:ethier-diel-char-curr-ff}
\end{figure}

As presented in~\eqref{eq:eigenCurrent1}, once the eigenvector~$\M{a}_n$ is obtained from~\eqref{eq:Seig}, the evaluation of other characteristic quantities is straightforward. A sample calculation of characteristic current density and characteristic directivity patterns is presented in Fig.~\ref{fig:ethier-diel-char-curr-ff} for the dominant mode of PIFA~\#3. The low-frequency figure shows a formation of an electric dipole-like radiation backed by the ground plane, still having considerable back radiation. At resonance of the dominant mode, the radiation diagram resembles the one used on fed PIFAs. The expansion of the current density into substructure characteristic modes on fed single resonance PIFA would be dominated by this mode.

\subsection{PIFA Mounted on an Electrically Large Platform}

The previous example demonstrates the quantitative effect of dielectric substrates on modal characteristics of a PIFA, which can, to a certain extent, be anticipated from general rules of antenna design, \ie{}, shifting of resonances to lower frequencies through the use of dielectric loading. In this final example, however, we consider the modal characteristics of a complex system for which limited engineering intuition can be applied \textit{a priori}.

\begin{figure}
    \centering
    \includegraphics[width=\columnwidth]{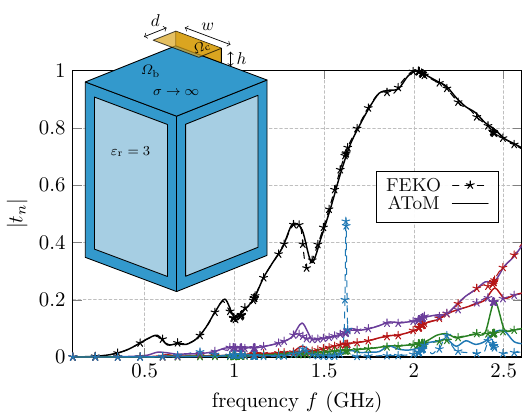}
    \caption{The first five substructure characteristic modes of a PIFA mounted on a frame. The frame has dimensions $10 \times 10 \times 15$\,cm$^3$ and is made of PEC and has a width of 1\ cm. The top cover is also made of PEC and forms a ground plane for a PIF. The PIFA has dimensions of $w=5$\,cm, $d=2.5$\,cm, $h=1.25$\,cm, and is the only controllable region in the model. The frame is filled with dielectrics of relative permittivity $\varepsilon_\T{r}=3$. The meaning of the solid and dashed lines is the same as in Fig.~\ref{fig:ethier-diel}.}
    \label{fig:cubesat}
\end{figure}

The substructure characteristic modes of a PIFA mounted on an electrically large platform are evaluated in Fig.~\ref{fig:cubesat}. The frame of dimensions $10 \times 10 \times 15$\,cm$^3$ (approximately 1.5U CubeSat format~\cite{cubeSat}) is formed by a PEC strip of width 1\,cm, and is galvanically connected with the solid PEC top cover and left open on the underside. The interior is filled with a dielectric material with relative permittivity~$\varepsilon_\T{r}=3$. The PIFA of dimensions $w=5$\,cm, $d=2.5\,$cm, and $h=1.25\,$cm is mounted on the top cover of the frame and, similarly as in Fig.~\ref{fig:ethier-sub}, only this antenna region is considered controllable for modal analysis, see the inset in Fig.~\ref{fig:cubesat}. The model is lossless and treated with an iterative algorithm that employs surface equivalence method-of-moment from Altair FEKO~\cite{feko}. The traces were adaptively refined to obtain smooth curves with relatively few samples. Numerical solver AToM~\cite{atom} was employed to independently verify the traces in a similar manner as in Fig.~\ref{fig:ethier-diel}.

The spectrum shown in Fig.~\ref{fig:cubesat} is dominated by only one mode, which agrees well with the similar arrangement in Fig.~\ref{fig:ethier-sub}. Considering the 20\% relative size reduction of PIFA in Fig.~\ref{fig:cubesat} as compared with Fig.~\ref{fig:ethier-sub}, the dominant mode resonates at comparable frequency. The background structure contains a large dielectric block with many internal resonances (approximately 80 cavity resonances between 0.1\,GHz and 2.6\,GHz). Most of these resonances are filtered out by the substructure formulation~\eqref{eq:Seig}. The only exception is the mode having an abrupt increase of modal significance around 1.6\,GHz.  The reason is that this peak, similar to the wavy behavior of the red, green, and blue traces around 2.5\,GHz, is a numerical artifact caused by a sensitive and imperfect cancellation between eigenvalues of $\M{S}$ and $\M{S}_\T{b}$ matrices, \cf{},~\eqref{eq:Seig}. This happens when the background region dominates the EM behavior of the entire structure, \ie{}, when the spectra of $\M{S}$ and $\M{S}_\T{b}$ are similar. To be more specific and taking the explanation from Sec.~\ref{sec:IVa}, the sharp blue peak results from an imperfect cancellation of two eigenvalues~$t_n$ (in the complex plane) which are almost identical for the full case and the background. In contrast, the dominant substructure mode represented by the black line in Fig.~\ref{fig:cubesat} results from a mode that only exists on the full structure and has no pair in the spectrum of the background. This numerical issue was not observed in the previous examples where the electrical size of the background was not sufficient to overshadow the scattering behavior of the controllable region.

\section{Discussion}
\label{sec:discu}

The theoretical developments and set of examples provided in previous sections thoroughly demonstrate that the scattering formulation of substructure characteristic modes~\eqref{eq:Seig} is able to reproduce cases treated using classical formulation~\cite{Ethier+McNamara2012}, while also allowing for the analysis of arbitrarily complex material distributions without modifying the evaluation procedure. 

\begin{figure}
    \centering
    \includegraphics[width=\columnwidth]{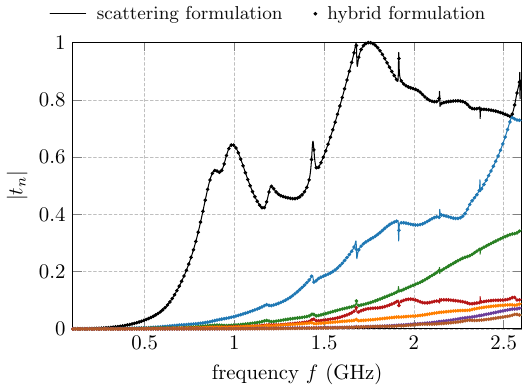}
    \caption{\TB{Modal significances~$|t_n|$ computed for the same setup as in Fig.~\ref{fig:hybridCMs} but with a sphere made of bi-isotropic medium with relative permittivity~$\varepsilon_\T{r} = 4$, relative permeability~$\mu_\T{r} = 1$, electro-magnetic susceptibility~$\chi^\T{em} = 1 + \T{j}$ and magneto-electric susceptibility~$\chi^\T{me} = \left( \chi^\T{em} \right)^\T{H}$. The presence of the real part of susceptibility~$\chi^\T{em}$ makes the system non-reciprocal. The impedance-based method relies on~\eqref{eq:CMsubMoM} and~\eqref{eq:Zhybrid}, while the scattering-based formulation relies on~\eqref{eq:Seig}.}}
    \label{fig:hybridCMsBiisotropic}
\end{figure}

\TB{As an example of treating a complex material distribution, ~Fig.~\ref{fig:hybridCMsBiisotropic} presents characteristic data for a setup similar to Fig.~\ref{fig:hybridCMs} but with a bi-isotropic medium described by constitutive relations}
\begin{equation}
\begin{aligned}
    \V{D} &= \varepsilon_0 \varepsilon_\T{r} \V{E} + c_0^{-1}\chi^\T{em} \V{H} \\
    \V{B} &= c_0^{-1}\chi^\T{me} \V{E} + \mu_0 \mu_\T{r} \V{H} 
\end{aligned},
\end{equation}
\TB{where $\varepsilon_0$, $\mu_0$, and $c_0$ are the vacuum permittivity, permeability, and speed of light, respectively.  The evaluation of the characteristic spectrum solely from integral equation solvers presents a considerable difficulty for the use of the bi-isotropic Green's function or the combined magneto-electric volume formulation. In contrast, the scattering-based formulation relying on~\eqref{eq:Seig} only demands the use of any electromagnetic solver that can solve two scattering problems.  Here the problem is solved using the scattering formulation \eqref{eq:Seig} and the hybrid MoM \& T-matrix method described in Sec.~\ref{sec:TMoMhybrid} as a benchmark.  Note that the latter does not represent a fully MoM-based implementation of this problem, as the authors are unaware of any such implementations are available for this class of problem.}

\subsection{Impact on Antenna Design}
With the exception of the PIFA mounted on conductive frame example shown in Fig.~\ref{fig:cubesat}, all examples are based on the same PIFA-like controllable region in the presence of varying backgrounds.  Substantial differences between the resulting modal characteristics in each example clearly illuminate the high potential impact of background objects on modal performance.  This is particularly noticeable in Fig.~\ref{fig:ethier-diel}, where the inclusion of thin dielectric support layers leads to non-negligible changes in modal significance.  If outputs from characteristic mode analyses are used in the design of antennas, it is therefore critical to fully model any background objects, including dielectrics, rather than using simplified background models and small perturbation approximations.  The approach presented in this manuscript facilitates this rigorous analysis.

\subsection{Generalization to Other Substructure Problems}
\begin{figure}%
    \centering
    \includegraphics[width=8.9cm]{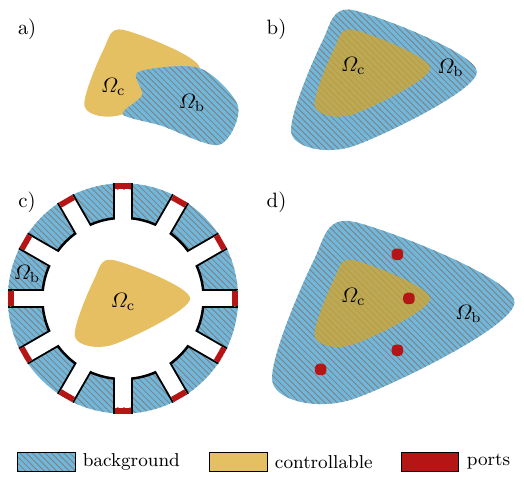}
    \caption{Illustrations of different setups to be considered for decomposition via substructure characteristic modes. (a) Classical arrangement with non-overlapping (possibly connected) regions. (b) Region~$\regc$ partly or fully overlapping with background region~$\regu$. (c) Region~$\regu$ in a cavity and $\regc$ also contains waveguide ports. (d) Region $\regc$ overlapping with background material in the presence of ports. Regions $\regc$ and $\regu$ are composed of arbitrary material distributions.}
\label{fig:subCMgeos}
\end{figure}

To further investigate the potential of the scattering formulation, it is worth considering the scenarios depicted in Fig.~\ref{fig:subCMgeos}. All the cases treated in Sections~\ref{sec:substr},~\ref{S:MoMsubstructure} and~\ref{sec:extSolver} solely dealt with panel~(a), in which controllable and background regions might share a boundary but are otherwise disjoint. 

The first generalization is shown in panel (b), where controllable and background regions share the same volumes yet have different material properties. The meaning of such a situation can be understood from the volume equivalence principle. When building an equivalent description of a given scattering scenario, one can choose which part of polarization belongs to equivalent sources and which belongs to a background. In such a case, the controllable region is not a physical structure but a contrast between two material distributions. In principle, this scenario can be approached by classical treatment using a partial background Green’s function, however, the computation overhead will be considerable. On the other hand, within the scattering formulation, the problem from panel~(b) is treated exactly like the problem from panel~(a), \ie{}, by separately evaluating scattering from the entire system and the background.

The second generalization is the addition of ports, a situation depicted in panel~(c). A possible scenario might be a substructure problem of Section~\ref{sec:substr}, where a port is added to controllable degrees of freedom\footnote{If the port is considered part of the background, the system will be lossy, which is not considered in this paper.}. Although proposals exist for evaluating characteristic modes on antennas loaded by ports~\cite{2021_Deng_AWPL}, we stress here that the ports present no qualitative change for the scattering formulation. Furthermore, the scattering formulation suggests that substructure characteristic modes can be evaluated even for circuits~\cite{Pozar2005}. For example, the scattering matrix~$\M{S}_\bSub$ might represent a circuit without a particular region, which is considered the controllable part. 

Lastly, panel~(d) of Fig.~\ref{fig:subCMgeos} combines all preceding scenarios, ports still belonging to the controllable degrees of freedom. A possible application of this scenario is the analysis and optimization of minimum scattering antenna~\cite{Kahn_Kurss_1965_MSA} by inspecting characteristic modes and their significance~\cite{Rogers_1986_MSA}, or techniques involving generalized scattering matrix~\cite{Kim_etal_2013_MSA, Alian_2023_MSA}. In such a case, the substructure characteristic modes are not reachable via the Schur complement method~\cite{Ethier+McNamara2012}, and their formulation will be challenging if background Green’s function is employed. Yet, the scattering formulation stays always the same, solely demanding to evaluate scattering matrices~$\M{S}$ and~$\M{S}_\bSub$ for a situation with and without the controllable part of the structure.

\begin{figure}%
    \centering
    \includegraphics[width=8.9cm]{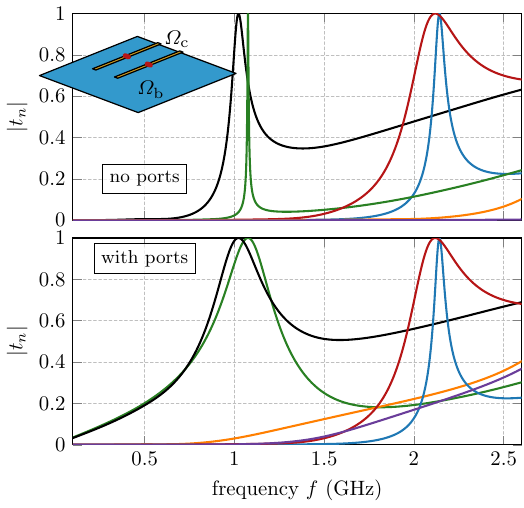}
    \caption{Modal significances~$\left| t_n \right|$ computed for the two parallel strips (controllable part) in the presence of a ground plane (background). The strips are potentially connected to lumped ports (controllable part). The strips have length~$l \approx 128 \, \T{mm}$ and width~$l/20$. The distance between the dipole centers is $l/3$. The ground plane is a square of edge~$3l/2$ and is positioned at the distance~$l/5$ below the strips. All materials are perfect electric conductors.}
\label{fig:subCMCaseC}
\end{figure}

To provide a particular example of scenarios from Fig.~\ref{fig:subCMgeos}c and Fig.~\ref{fig:subCMgeos}d, consider two co-planar strips lying above a finite ground plane, see the inset of Fig.~\ref{fig:subCMCaseC}. The strips are in their centers connected to delta-gap ports with impedance~$Z_0 = 73 \, \T{\Omega}$. The ground plane is considered as a background. The modal significances of substructure characteristic modes are shown in Fig.~\ref{fig:subCMCaseC}. The figure shows the case with ports being part of controllable degrees of freedom (corresponding to Fig.~\ref{fig:subCMgeos}c) as well as the case when ports are missing (corresponding to Fig.~\ref{fig:subCMgeos}b). It can be seen that the presence of ports has a considerable effect on the modal spectrum, particularly for modes with high current density at the location of the ports (green and black traces), which are effectively broadened by the inclusion of the ports. 
Two modes with zero current density at the position of the ports (red and blue) are unchanged between the models with and without ports.
It is also important to notice that far fields attached to the case of Fig.~\ref{fig:subCMgeos}c or Fig.~\ref{fig:subCMgeos}d are orthogonal only when waves exiting the ports are also taken as a part of the far field. The sole radiation diagrams are not orthogonal as they are in the cases of Fig.~\ref{fig:subCMgeos}a and Fig.~\ref{fig:subCMgeos}b.

\subsection{Numerical Accuracy and Computational cost}

The scattering formulation offers the ability to use arbitrary numerical tools when evaluating substructure characteristic modes.  In this way, it allows for the judicious selection of numerical tools that can accurately and efficiently evaluate scattering from the particular structure and background being studied.  As discussed in Section~\ref{sec:comp-comp}, certain problems benefit from the use of particular numerical methods, including situations where integral equations (and by proxy the existing Schur-complement approach to substructure modes) are the optimal choice for efficiency and accuracy.

Here, we reiterate that the scattering formulation opens the door for the analysis of substructure characteristic modes using other numerical methods and that the primary benefit of this extension is the ability to solve problems that are difficult or impossible to efficiently or accurately study using integral equations.  Regardless of which method or numerical tool is selected, it is important to keep in mind that, ultimately, the accuracy of modal data relies on the accuracy of the underlying numerical modeling, be it scattering simulations or the construction of integral operators.

\section{Conclusion}
\label{sec:conclu}

The method proposed in this paper extends the scattering-based formulation of characteristic modes to general substructure problems.  Such problems were previously limited to cases where numerical or analytical problem-specific (non-free-space) Green’s functions are known, typically by way of Schur complement methods based on the method of moments.  Using the proposed extension of scattering-based characteristic mode analysis, this requirement is lifted, and characteristic modes for all linear problems with arbitrary structure/substructure designations can be analyzed.  

The examples throughout the paper demonstrate the flexibility of the method through its application to a family of problems based on a PIFA-like antenna above a finite ground plane.  The resulting data demonstrate equivalence between \TB{the modes from } the proposed method \TB{with those from } impedance-based formulations \TB{for problems where both are applicable}, and highlight the generality of the approach to variations involving large dielectric background media, infinite ground planes, and finite dielectric regions.   Further discussion regarding its application to problems involving wave ports and acceleration using iterative algorithms points to several areas for continued research, with the general direction aiming toward the fast and efficient characteristic mode analysis of antennas and scatterers in highly complex, arbitrary material environments.

\appendices

\section{Matrix-Free Evaluation of Substructure Characteristic Modes}
\label{S:ItCMsub}

Characteristic modes can be evaluated efficiently using an iterative matrix-free algorithm~\cite{Lundgren+etal2023}. Instead of having full, explicit knowledge of a matrix being decomposed, these algorithms rely on knowing the result of a matrix-vector multiplication of that matrix with an arbitrary vector.  In the case of characteristic modes, the matrix itself is a scattering operator, while the matrix-vector multiplication represents the solution to a scattering problem for a particular excitation. Nevertheless, algorithms such as Arnoldi iteration~\cite{Golub+Loan2013} used in~\cite{Lundgren+etal2023} are not suitable for generalized eigenvalue problems, such as~\eqref{eq:Seig}. Therefore, to evaluate substructure characteristic modes in a matrix-free fashion, the formulations~\eqref{eq:Seigalt},~\eqref{eq:TeigA} and~\eqref{eq:Teigf}  are used. Furthermore, it is important to realize that only results of multiplications~$\M{T} \M{x}, \M{T}_\bSub \M{x}, \M{S} \M{x}, \M{S}_\bSub \M{x}$ are accessible via general purpose electromagnetic solvers. When adapting the algorithms described in~\cite{Lundgren+etal2023}, the following changes in the desired matrix-vector products are made
\begin{equation}
\begin{aligned}
    \M{S}_\bSub^{\T{H}}\M{S}\M{a}_n &= \left( \M{S}_\bSub \left(\M{S}\M{a}_n \right)^* \right)^* \\
    (2\M{T}_\bSub^{\T{H}}\M{T}+\M{T}_\bSub^{\T{H}}+\M{T})\M{a}_n &= \left( \M{T}_\bSub \left(\M{a}_n + 2 \M{T}\M{a}_n \right)^* \right)^* + \M{T}\M{a}_n
\end{aligned}    
\end{equation}
where it was assumed that scattering, as well as transition matrices, are symmetric and where only formulations involving excitation vectors~$\M{a}_n$ are shown for brevity.

\begin{algorithm}
\caption{Matrix-free algorithm}\label{alg:T}
\begin{spacing}{1.15}
\begin{algorithmic}[1]
\State\label{state:init-m} $m = 0$
\State\label{state:init-ex} $\M{a}_\bSub \gets \T{rand}$
\While{stopping criteria are not met}\label{state:while}
\State \label{state:norm}$\M{a}_m \gets \M{a}_m/|\M{a}_m|$
\State \label{state:MGS} modified Gram-Schmidt procedure over $\{\M{a}_m\}$
\State\label{state:fullwave1} $\M{f}_{1m} \gets \M{M} \M{a}_m$
\State\label{state:fullwave2B} $\widehat{\M{a}} \gets \M{P} \left(\M{a}_m + 2 \M{f}_{1m} \right)^*$
\State\label{state:fullwave2} $\widehat{\M{f}}_{2m} \gets \M{M}_\bSub \widehat{\M{a}}$
\State\label{state:fullwave3} $\M{f}_m \gets \M{f}_{1m} + \M{P} \widehat{\M{f}}_{2m}^\ast$
\State\label{state:update-t} $\M{A}_m \gets \sum_{p\leq m} \M{f}_p\M{a}_p^{\herm}$
\State\label{state:est-t} $\{t_n\}_m\gets \T{eig}(\M{A}_m)$
\State\label{state:update-p} $\M{P}_m \gets \sum_{p\leq m} 
\M{a}_p\M{a}_p^{\herm}$
\State\label{state:update-a} $\M{a}_{m+1} \gets \M{f}_m - \M{P}_m\M{f}_m$
\State $m \gets m+1$
\EndWhile
\end{algorithmic}
\end{spacing}
\end{algorithm}

An example of the procedure used to estimate substructure characteristic modes using transition matrices in matrix-free manner is sketched in Algorithm~\ref{alg:T}, where~$\M{M} = \M{T}$, $\M{M}_\bSub = \M{T}_\bSub$, $\M{A} = 2\M{T}_\bSub^{\T{H}}\M{T}+\M{T}_\bSub^{\T{H}}+\M{T}$ abbreviates the matrix to be decomposed, and $\M{P} = \M{1}$ is an identity matrix. The algorithm is stopped when the magnitude~$\left|\M{a}_{m+1}\right|$ is sufficiently small or relative changes in estimated eigenvalues~$t_n$ are sufficiently small. Steps no.~5  and~6 are the solutions to scattering problems involving full structure and background, respectively, and can be obtained from any full-wave electromagnetic solver. The modified Gram–Schmidt procedure is used in Algorithm~\ref{alg:T} to assure its stability~\cite{Golub+Loan2013}. Algorithm~1 is, for simplicity, presented using matrices $\M{A}_m$ and $\M{P}_m$. However, it is important to note that storing these as full matrices is not required for evaluation of the eigenvalues~\cite[Ch. 10]{Golub+Loan2013}.

Another possibility to evaluate substructure characteristic modes is to employ scattering dyadic matrices, accessible with arbitrary electromagnetic solver~\cite{Capek+etal2023a}. In this case, the matrices $\nS$ and $\nS_\bSub$ are defined as in~\cite[eq.~(21)]{Capek+etal2023a} and denoted here as scattering and background scattering dyadic matrices, respectively. The matrices are not transposed symmetric~\cite[eq.~(4)]{Capek+etal2023a}, so the Algorithm~\ref{alg:T} must be modified by settings~$\M{M} = \nS$,~$\M{M}_\bSub = \nS_\bSub$, and with matrix~$\M{P}$ being indexing matrix containing zeros except positions~$P_{pq}$ and~$P_{qp}$ where pairs of the quadrature points are mapped as
\begin{equation}
    P_{pq} = P_{qp} = 1 \Longleftrightarrow \hat{\V{r}}_p = - \hat{\V{r}}_q.
\end{equation}
In addition, matrix~$\M{P}$ is further modified by flipping the $\pm$~sign for respective positions where the quadrature points in $\vartheta$-polarization block of the dyadic lie on $\pm z$ axis, and for all entries corresponding to $\varphi$-polarization block except of points lying on $\pm z$~axis.

\section{Equivalence Between MoM-based and Scattering-Based Substructure Characteristic Modes}\label{S:SS0MoMeq}

The modified transition matrix~\eqref{eq:Ttil} resembles the transition matrix-based expression in~\cite{Gustafsson+etal2022a} with the difference that the spherical wave matrix $\widetilde{\M{U}}$ is complex-valued.
Rewriting~\eqref{eq:Ttil} using block matrices~\eqref{eq:SubMoM} produces
\begin{equation}
\wtM{T}=	
\M{U}_1
\begin{bmatrix}
		-\M{Z}_{\bSub\bSub}^{-1}\M{Z}_{\bSub\cSub}\widetilde{\M{Z}}^{-1}\M{Z}_{\bSub\cSub}^{\herm}\M{Z}_{\bSub\bSub}^{-\herm} & 
		\M{Z}_{\bSub\bSub}^{-1}\M{Z}_{\bSub\cSub}\widetilde{\M{Z}}^{-1} \\[5pt]
		\widetilde{\M{Z}}^{-1}\M{Z}_{\bSub\cSub}^{\herm}\M{Z}_{\bSub\bSub}^{-\herm}
		& -\widetilde{\M{Z}}^{-1}
	\end{bmatrix}	
 \M{U}_1^{\trans}
\label{eq:SWMoMeig}
\end{equation}
which partly resembles a block inversion of the MoM matrix~\eqref{eq:SubMoM} except for some Hermitian transposes. 

To express the transition matrices of the composite object~$\reg$ and background object~$\regu$ in MoM system matrices, we use~\cite{Gustafsson+etal2022a} 
\begin{equation}
	\M{T} = -\M{U}_1\M{Z}^{-1}\M{U}_1^{\trans}
    \quad\text{and }
	\M{T}_{\bSub} 
	=-
		\M{U}_1
	\begin{bmatrix}
		\M{Z}_{\bSub\bSub}^{-1} & \M{0}\\
		\M{0} & \M{0} 
	\end{bmatrix}
		\M{U}_1^{\trans}.
\label{eq:Ttot}
\end{equation}
Substituting these T-matrices into~\eqref{eq:Teigf} and using block matrix inversion together with algebraic manipulations outlined below we realize that~\eqref{eq:CMsubMoM},~\eqref{eq:CMeigUZU},~\eqref{eq:Teigf} are all identical.

The derivation starts with reformulation of~\eqref{eq:Teigf} in MoM matrices
\begin{multline}
    2\M{T}\M{T}^{\T{H}}_\bSub+\M{T}_\bSub^{\T{H}}+\M{T}
    \\ =
    \M{U}_1 \M{Z}^{-1} \left( 2
 \M{R}
  \begin{bmatrix}
    \M{Z}_{\bSub\bSub}^{-\herm} & \M{0}\\
    \M{0} & \M{0} 
\end{bmatrix}
	-\M{Z}
 \begin{bmatrix}
    \M{Z}_{\bSub\bSub}^{-\herm} & \M{0}\\
    \M{0} & \M{0} 
\end{bmatrix}
- \M{1} \right) \M{U}_1^{\trans},
\end{multline} 
where~$\M{R} = \M{U}_1^\trans \M{U}_1$ has been used~\cite{Tayli+etal2018}. Realizing further that~$2 \M{R} = \M{Z} + \M{Z}^\herm$ for lossless scatterers, the relation simplifies to
\begin{multline}
2\M{T}\M{T}^{\T{H}}_\bSub+\M{T}_\bSub^{\T{H}}+\M{T} = 
\M{U}_1\M{Z}^{-1}\left(\M{Z}^{\herm}
		\begin{bmatrix}
		\M{Z}_{\bSub\bSub}^{-\herm} & \M{0}\\
		\M{0} & \M{0} 
	\end{bmatrix}	
	-\M{1}
	\right)	\M{U}_1^{\trans} \\
 = \M{U}_1 \M{Z}^{-1}
 	\begin{bmatrix}
		\M{0} &  \M{0}\\
		\M{Z}_{\bSub\cSub}^{\herm}\M{Z}_{\bSub\bSub}^{-\herm} & -\M{1} 
	\end{bmatrix} \M{U}_1^{\trans}.
\label{eq:}
\end{multline}
The final step is the use of block matrix inversion 
\begin{equation}
\M{Z}^{-1}
=	\begin{bmatrix}
		\M{Z}_{\bSub\bSub}^{-1} + \M{Z}_{\bSub\bSub}^{-1}\M{Z}_{\bSub\cSub}\widetilde{\M{Z}}^{-1}\M{Z}_{\cSub\bSub}\M{Z}_{\bSub\bSub}^{-1} & -\M{Z}_{\bSub\bSub}^{-1}\M{Z}_{\bSub\cSub}\widetilde{\M{Z}}^{-1} \\[5pt]
	-\widetilde{\M{Z}}^{-1}\M{Z}_{\cSub\bSub}\M{Z}_{\bSub\bSub}^{-1} & \widetilde{\M{Z}}^{-1} 
	\end{bmatrix}
\label{eq:}
\end{equation}
which is identical to $\wtM{T}$ in~\eqref{eq:SWMoMeig}
\begin{equation}
    2\M{T}\M{T}^{\T{H}}_\bSub+\M{T}_\bSub^{\T{H}}+\M{T} = \wtM{T}
\end{equation}
proving that~\eqref{eq:CMsubMoM},~\eqref{eq:CMeigUZU},~\eqref{eq:Teigf} are the same.

Notice that the above derivation demanded lossless scatterer. For lossy cases, matrix~$\widetilde{\M{R}}$ differs from the radiation operator $\widetilde{\M{U}}_1^{\herm}\widetilde{\M{U}}_1$. For the special case with lossless background, the scattering-based formulation of characteristic modes is identical to the $\widetilde{\M{Z}}$-formulation on the right in~\eqref{eq:CMsubMoM}.


\begin{IEEEbiography}[{\includegraphics[width=1in,height=1.25in,clip,keepaspectratio]{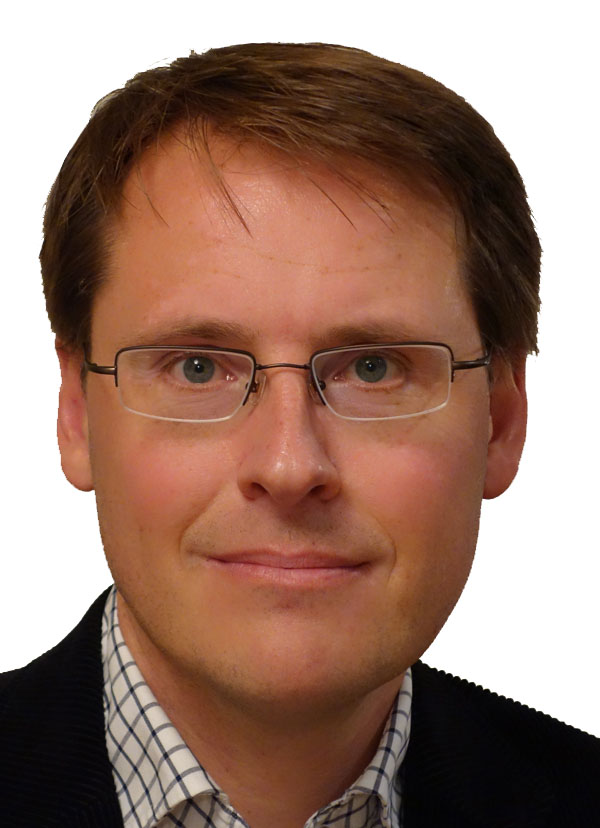}}]{Mats Gustafsson}
received the M.Sc. degree in Engineering Physics 1994, the Ph.D. degree in Electromagnetic Theory 2000, was appointed Docent 2005, and Professor of Electromagnetic Theory 2011, all from Lund University, Sweden.

He co-founded the company Phase holographic imaging AB in 2004. His research interests are in scattering and antenna theory and inverse scattering and imaging. He has written over 100 peer reviewed journal papers and over 100 conference papers. Prof. Gustafsson received the IEEE Schelkunoff Transactions Prize Paper Award 2010, the IEEE Uslenghi Letters Prize Paper Award 2019, and best paper awards at EuCAP 2007 and 2013. He served as an IEEE AP-S Distinguished Lecturer for 2013-15.
\end{IEEEbiography}

\begin{IEEEbiography}[{\includegraphics[width=1in,height=1.25in,clip,keepaspectratio]{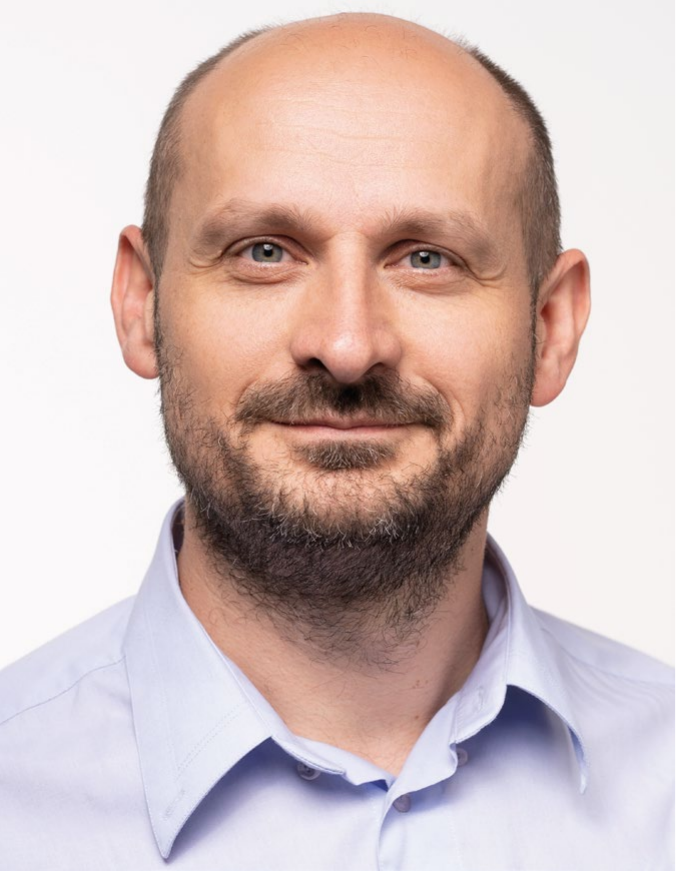}}]{Lukas Jelinek} was born in Czech Republic in 1980. He received his Ph.D. degree from the Czech Technical University in Prague, Czech Republic, in 2006. In 2015 he was appointed Associate Professor at the Department of Electromagnetic Field at the same university.

His research interests include wave propagation in complex media, electromagnetic field theory, metamaterials, numerical techniques, and optimization.
\end{IEEEbiography}

\begin{IEEEbiography}[{\includegraphics[width=1in,height=1.25in,clip,keepaspectratio]{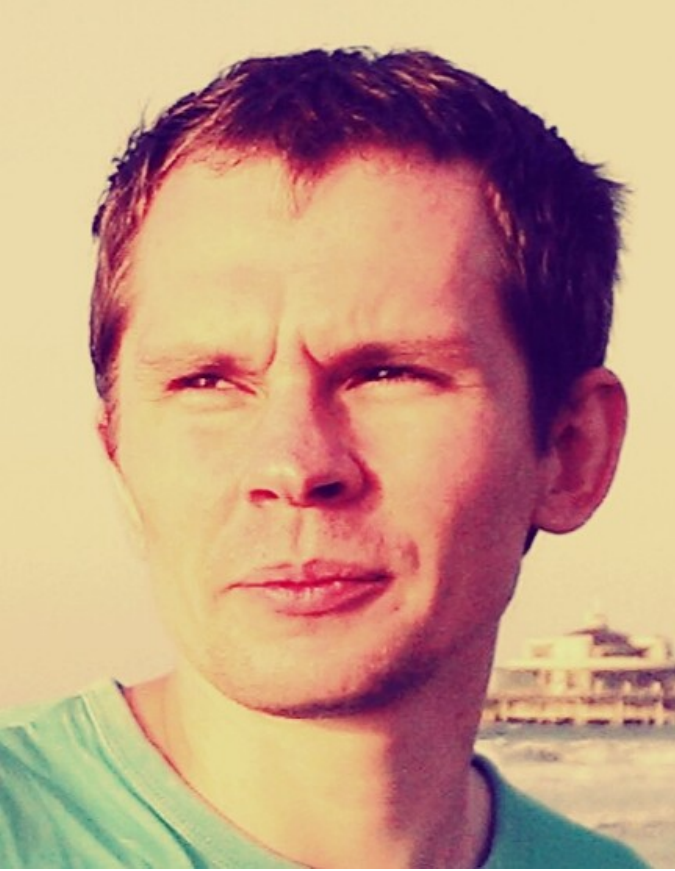}}]{Miloslav Capek}
(M'14, SM'17) received the M.Sc. degree in Electrical Engineering 2009, the Ph.D. degree in 2014, and was appointed a Full Professor in 2023, all from the Czech Technical University in Prague, Czech Republic.
	
He leads the development of the AToM (Antenna Toolbox for Matlab) package. His research interests include electromagnetic theory, electrically small antennas, antenna design, numerical techniques, and optimization. He authored or co-authored over 160~journal and conference papers.

Dr. Capek is the Associate Editor of IET Microwaves, Antennas \& Propagation. He was a regional delegate of EurAAP between 2015 and 2020. He received the IEEE Antennas and Propagation Edward E. Altshuler Prize Paper Award~2023.
\end{IEEEbiography}

\begin{IEEEbiography}[{\includegraphics[width=1in,height=1.25in,clip,keepaspectratio]{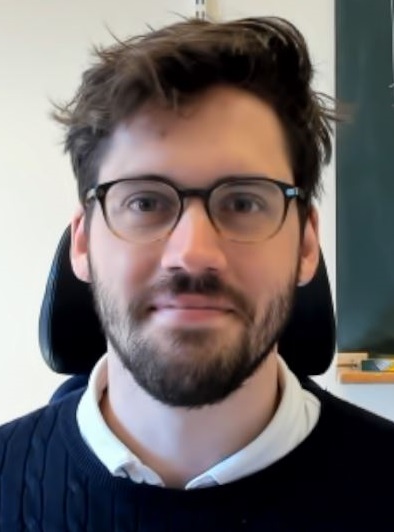}}]{Johan Lundgren}
(M'22) is an assistant professor at Lund University. he received his M.Sc degree in engineering physics 2016 and Ph.D. degree in Electromagnetic Theory in 2021 all from Lund University, Sweden. 

His research interests are in electromagnetic scattering, wave propagation, computational electromagnetics, functional structures, meta-materials, inverse scattering problems, imaging, and measurement techniques.

\end{IEEEbiography}

\begin{IEEEbiography}[{\includegraphics[width=1in,height=1.25in,clip,keepaspectratio]{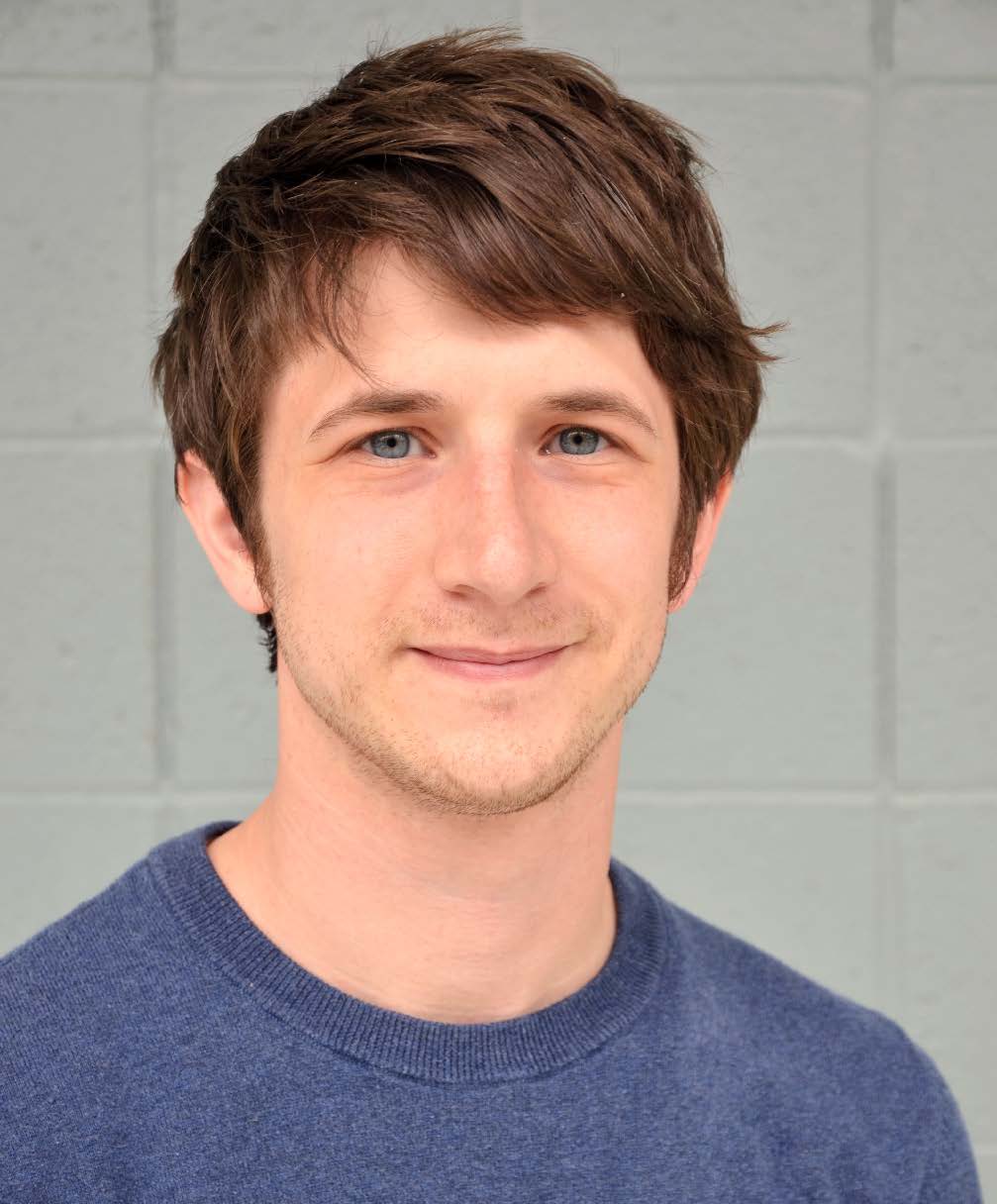}}]{Kurt Schab}
(S'09, M'16) is an Assistant Professor of Electrical Engineering at Santa Clara University, Santa Clara, CA USA. He received the B.S. degree in electrical engineering and physics from Portland State University in 2011 and the M.S. and Ph.D. degrees in electrical engineering from the University of Illinois at Urbana-Champaign in 2013 and 2016, respectively.  From 2016 to 2018 he was a Postdoctoral Research Scholar at North Carolina State University in Raleigh, North Carolina.  His research focuses on the intersection of numerical methods, electromagnetic theory, and antenna design.  
\end{IEEEbiography}


\begin{thebibliography}{10}
	\providecommand{\url}[1]{#1}
	\csname url@samestyle\endcsname
	\providecommand{\newblock}{\relax}
	\providecommand{\bibinfo}[2]{#2}
	\providecommand{\BIBentrySTDinterwordspacing}{\spaceskip=0pt\relax}
	\providecommand{\BIBentryALTinterwordstretchfactor}{4}
	\providecommand{\BIBentryALTinterwordspacing}{\spaceskip=\fontdimen2\font plus
		\BIBentryALTinterwordstretchfactor\fontdimen3\font minus
		\fontdimen4\font\relax}
	\providecommand{\BIBforeignlanguage}[2]{{%
			\expandafter\ifx\csname l@#1\endcsname\relax
			\typeout{** WARNING: IEEEtran.bst: No hyphenation pattern has been}%
			\typeout{** loaded for the language `#1'. Using the pattern for}%
			\typeout{** the default language instead.}%
			\else
			\language=\csname l@#1\endcsname
			\fi
			#2}}
	\providecommand{\BIBdecl}{\relax}
	\BIBdecl
	
	\bibitem{Montgomery+Dicke+Purcell1948}
	C.~G. Montgomery, R.~H. Dicke, and E.~M. Purcell, \emph{Principles of Microwave
		Circuits}.\hskip 1em plus 0.5em minus 0.4em\relax New York, NY: McGraw-Hill,
	1948.
	
	\bibitem{Garbacz_1965_TCM}
	R.~Garbacz, ``Modal expansions for resonance scattering phenomena,''
	\emph{Proc. IEEE}, vol.~53, no.~8, pp. 856--864, Aug. 1965.
	
	\bibitem{Harrington+Mautz1971}
	R.~F. Harrington and J.~R. Mautz, ``Theory of characteristic modes for
	conducting bodies,'' \emph{IEEE Trans. Antennas Propag.}, vol.~19, no.~5, pp.
	622--628, 1971.
	
	\bibitem{lau2022characteristic}
	B.~K. Lau, M.~Capek, and A.~M. Hassan, ``Characteristic modes: Progress,
	overview, and emerging topics,'' \emph{IEEE Antennas and Propagation
		Magazine}, vol.~64, no.~2, pp. 14--22, 2022.
	
	\bibitem{Gustafsson+etal2022a}
	M.~Gustafsson, L.~Jelinek, K.~Schab, and M.~Capek, ``Unified theory of
	characteristic modes: Part {I}--fundamentals,'' \emph{IEEE Trans. Antennas
		Propag.}, vol.~70, no.~12, pp. 11\,801--11\,813, 2022.
	
	\bibitem{Gustafsson+etal2022b}
	------, ``Unified theory of characteristic modes: Part {II}--tracking, losses,
	and {FEM} evaluation,'' \emph{IEEE Trans. Antennas Propag.}, vol.~70, no.~12,
	pp. 11\,814--11\,824, 2022.
	
	\bibitem{Capek+etal2023a}
	M.~Capek, J.~Lundgren, M.~Gustafsson, K.~Schab, and L.~Jelinek,
	``Characteristic mode decomposition using the scattering dyadic in arbitrary
	full-wave solvers,'' \emph{IEEE Trans. Antennas Propag.}, vol.~71, no.~1, pp.
	830--839, 2023.
	
	\bibitem{Lundgren+etal2023}
	J.~Lundgren, K.~Schab, M.~Capek, M.~Gustafsson, and L.~Jelinek, ``Iterative
	calculation of characteristic modes using arbitrary full-wave solvers,''
	\emph{IEEE Antennas Wireless Propag. Lett.}, vol.~22, no.~4, pp. 799--803,
	2023.
	
	\bibitem{Ethier+McNamara2012}
	J.~Ethier and D.~McNamara, ``Sub-structure characteristic mode concept for
	antenna shape synthesis,'' \emph{Electronics letters}, vol.~48, no.~9, p.~1,
	2012.
	
	\bibitem{alroughani2013appraisal}
	H.~Alroughani, ``An appraisal of the characteristic modes of composite
	objects,'' Master's thesis, University of Ottawa (Canada), 2013.
	
	\bibitem{Angiulli+etal2000}
	G.~Angiulli, G.~Amendola, and G.~Di~Massa, ``Application of characteristic
	modes to the analysis of scattering from microstrip antennas,'' \emph{J.
		Electromagnet. Waves Appl.}, vol.~14, no.~8, pp. 1063--1081, 2000.
	
	\bibitem{2023_Zhao_TAP}
	R.~Zhao, Y.~Lu, G.~S. Cheng, W.~Zhu, J.~Hu, and H.~Bagci, ``Sub-structure
	characteristic mode analysis of microstrip antennas using a global multitrace
	formulation,'' \emph{IEEE Trans. Antennas Propag.}, vol.~71, no.~12, pp.
	10\,026--10\,031, 2023.
	
	\bibitem{Luomaniemi_CMA2021}
	R.~Luomaniemi, P.~Ylä-Oijala, A.~Lehtovuori, and V.~Viikari, ``Designing
	hand-immune handset antennas with adaptive excitation and characteristic
	modes,'' \emph{IEEE Trans. Antennas Propag.}, vol.~69, no.~7, pp. 3829--3839,
	2021.
	
	\bibitem{Parhami+etal1977}
	P.~Parhami, Y.~Rahmat-Samii, and R.~Mittra, ``Technique for calculating the
	radiation and scattering characteristics of antennas mounted on a finite
	ground plane,'' in \emph{Proceedings of the Institution of Electrical
		Engineers}, vol. 124, no.~11.\hskip 1em plus 0.5em minus 0.4em\relax IET,
	1977, pp. 1009--1016.
	
	\bibitem{cwik1989constructing}
	T.~Cwik, J.~Patterson, and T.~Lockhart, ``Constructing matrix {G}reen's
	functions for radiation and scattering problems,'' in \emph{Digest on
		Antennas and Propagation Society International Symposium}, 1989, pp. 586--589
	vol.2.
	
	\bibitem{dai2016characteristic}
	Q.~I. Dai, H.~Gan, W.~C. Chew, and C.-F. Wang, ``Characteristic mode analysis
	using {G}reen's function of arbitrary background,'' in \emph{2016 IEEE
		International Symposium on Antennas and Propagation (APSURSI)}.\hskip 1em
	plus 0.5em minus 0.4em\relax IEEE, 2016, pp. 423--424.
	
	\bibitem{alakhras2020sub}
	A.~Alakhras and D.~McNamara, ``Sub-structure characteristic mode computation
	utilising field-based {MM/GTD} hybrid methods,'' \emph{Journal of
		Electromagnetic Waves and Applications}, vol.~34, no.~13, pp. 1812--1821,
	2020.
	
	\bibitem{Harrington1968}
	R.~F. Harrington, \emph{Field Computation by Moment Methods}.\hskip 1em plus
	0.5em minus 0.4em\relax New York, NY: Macmillan, 1968.
	
	\bibitem{2014_Alroughani_ICEAA}
	H.~Alroughani, J.~L.~T. Ethier, and D.~A. McNamara, ``On the classification of
	characteristic modes, and the extension of sub-structure modes to include
	penetrable material,'' in \emph{2014 International Conference on
		Electromagnetics in Advanced Applications (ICEAA)}.\hskip 1em plus 0.5em
	minus 0.4em\relax IEEE, 2014.
	
	\bibitem{wu2019general}
	Q.~Wu, ``General metallic-dielectric structures: A characteristic mode analysis
	using volume-surface formulations,'' \emph{IEEE Antennas and Propagation
		Magazine}, vol.~61, no.~3, pp. 27--36, 2019.
	
	\bibitem{wu2019characteristic}
	------, ``Characteristic mode assisted design of dielectric resonator antennas
	with feedings,'' \emph{IEEE Trans. Antennas Propag.}, vol.~67, no.~8, pp.
	5294--5304, 2019.
	
	\bibitem{boyuan2021sie}
	M.~Boyuan, S.~Huang, D.~Yang, and J.~Pan, ``{SIE}-based substructure
	characteristic mode analysis on stacked dielectric resonator antennas,'' in
	\emph{2021 IEEE MTT-S International Microwave Workshop Series on Advanced
		Materials and Processes for RF and THz Applications (IMWS-AMP)}.\hskip 1em
	plus 0.5em minus 0.4em\relax IEEE, 2021, pp. 355--357.
	
	\bibitem{2021_Huang_TAP}
	S.~Huang, C.-F. Wang, J.~Pan, and D.~Yang, ``Accurate sub-structure
	characteristic mode analysis of dielectric resonator antennas with finite
	ground plan,'' \emph{IEEE Trans. Antennas Propag.}, vol.~69, no.~10, p.
	6930–6935, 2021.
	
	\bibitem{wu2022characteristic}
	Q.~Wu, ``Characteristic mode formulations of composite {PEC} and lossy
	dielectric objects,'' \emph{IEEE Antennas Wireless Propag. Lett.}, vol.~21,
	no.~8, pp. 1557--1561, 2022.
	
	\bibitem{guo2022novel}
	C.~Guo and Y.-C. Jiao, ``A novel radiated-power-based sub-structure
	characteristic mode formulation considering lossy background structures,'' in
	\emph{2022 International Applied Computational Electromagnetics Society
		Symposium (ACES-China)}.\hskip 1em plus 0.5em minus 0.4em\relax IEEE, 2022,
	pp. 1--2.
	
	\bibitem{2023_Huang_TAP}
	S.~Huang, C.-F. Wang, and M.-C. Tang, ``Generalized
	surface-integral-equation-based sub-structure characteristic-mode solution to
	composite objects,'' \emph{IEEE Trans. Antennas Propag.}, vol.~71, no.~3, p.
	2626–2639, 2023.
	
	\bibitem{guo2024substructure}
	C.~Guo, Y.-C. Jiao, and Y.~Ren, ``The substructure characteristic mode analysis
	for low-profile cavity-backed slot antennas,'' \emph{IEEE Trans. Antennas
		Propag.}, vol.~72, no.~8, pp. 6729--6734, 2024.
	
	\bibitem{2024_Guo_TAP}
	------, ``An alternative surface-integral-equation-based sub-structure
	characteristic mode formulation for lossy composite structures,'' \emph{IEEE
		Trans. Antennas Propag.}, vol.~72, no.~1, p. 707–719, 2024.
	
	\bibitem{Kristensson2016}
	G.~Kristensson, \emph{Scattering of Electromagnetic Waves by Obstacles}.\hskip
	1em plus 0.5em minus 0.4em\relax Edison, NJ: SciTech Publishing, an imprint
	of the IET, 2016.
	
	\bibitem{Pozar2005}
	D.~M. Pozar, \emph{Microwave Engineering}, 3rd~ed.\hskip 1em plus 0.5em minus
	0.4em\relax New York, NY: John Wiley \& Sons, 2005.
	
	\bibitem{Mishchenko+etal1996}
	M.~I. Mishchenko, L.~D. Travis, and D.~W. Mackowski, ``T-matrix computations of
	light scattering by nonspherical particles: {A} review,'' \emph{J. Quant.
		Spectrosc. Radiat. Transfer}, vol.~55, no.~5, pp. 535--575, 1996.
	
	\bibitem{capek2022computational}
	M.~Capek and K.~Schab, ``Computational aspects of characteristic mode
	decomposition: An overview,'' \emph{IEEE Antennas and Propagation Magazine},
	vol.~64, no.~2, pp. 23--31, 2022.
	
	\bibitem{Gustafsson+etal2023}
	\BIBentryALTinterwordspacing
	M.~Gustafsson, J.~Lundgren, K.~Schab, L.~Jelinek, and M.~Capek, ``Scattering
	matrix formulation for substructure characteristic mode analysis,'' in
	\emph{Proceedings of the XXXVth URSI General Assembly and Scientific
		Symposium (URSI GASS)}, 2023. [Online]. Available:
	\url{https://www.ursi.org/proceedings/procGA23/papers/2276.pdf}
	\BIBentrySTDinterwordspacing
	
	\bibitem{Song+Chew2001b}
	J.~Song and W.~C. Chew, ``Error analysis for the truncation of multipole
	expansion of vector {G}reen's functions,'' \emph{IEEE microwave and wireless
		components letters}, vol.~11, no.~7, pp. 311--313, 2001.
	
	\bibitem{Tayli+etal2018}
	D.~Tayli, M.~Capek, L.~Akrou, V.~Losenicky, L.~Jelinek, and M.~Gustafsson,
	``Accurate and efficient evaluation of characteristic modes,'' \emph{IEEE
		Trans. Antennas Propag.}, vol.~66, no.~12, pp. 7066--7075, 2018.
	
	\bibitem{Garbacz+Turpin1971}
	R.~J. Garbacz and R.~H. Turpin, ``A generalized expansion for radiated and
	scattered fields,'' \emph{IEEE Trans. Antennas Propag.}, vol.~19, no.~3, pp.
	348--358, 1971.
	
	\bibitem{Schab+etal2023}
	K.~Schab, F.~W. Chen, L.~Jelinek, M.~Capek, J.~Lundgren, and M.~Gustafsson,
	``Characteristic modes of frequency-selective surfaces and metasurfaces from
	{S}-parameter data,'' \emph{IEEE Trans. Antennas Propag.}, vol.~71, no.~12,
	pp. 9696--9706, 2023.
	
	\bibitem{schab2021sparsity}
	K.~Schab, ``Sparsity of radiating characteristic modes on infinite periodic
	structures,'' \emph{IEEE Trans. Antennas Propag.}, vol.~21, no.~2, pp.
	312--316, 2021.
	
	\bibitem{Chew+etal2001}
	W.~C. Chew, E.~Michielssen, J.~Song, and J.~Jin, \emph{Fast and Efficient
		Algorithms in Computational Electromagnetics}.\hskip 1em plus 0.5em minus
	0.4em\relax Artech House, Inc., 2001.
	
	\bibitem{Polimeridis+etal2014}
	A.~G. Polimeridis, J.~F. Villena, L.~Daniel, and J.~K. White, ``Stable
	{FFT-JVIE} solvers for fast analysis of highly inhomogeneous dielectric
	objects,'' \emph{Journal of Computational Physics}, vol. 269, pp. 280--296,
	2014.
	
	\bibitem{Wu+Wu2024}
	D.~Wu and Q.~Wu, ``A nested iterative solver for substructure {CMA} of
	composite metallic–dielectric objects,'' \emph{IEEE Antennas and Wireless
		Propagation Letters}, vol.~23, no.~5, pp. 1568--1572, 2024.
	
	\bibitem{Dai+etal2016}
	Q.~I. Dai, H.~Gan, C.~C. Weng, and C.-F. Wang, ``Characteristic mode analysis
	using {G}reen's function of arbitrary background,'' in \emph{2016 IEEE
		International Symposium on Antennas and Propagation (APSURSI)}.\hskip 1em
	plus 0.5em minus 0.4em\relax IEEE, 2016, pp. 423--424.
	
	\bibitem{Losenicky+etal2022}
	V.~Losenicky, L.~Jelinek, M.~Capek, and M.~Gustafsson, ``Method of moments and
	{T}-matrix hybrid,'' \emph{IEEE Trans. Antennas Propag.}, vol.~70, no.~5, pp.
	3560--3574, 2022.
	
	\bibitem{2022_Jelinek_APS}
	L.~Jelinek, M.~Gustafsson, K.~Schab, M.~Capek, and E.~Moreno, ``Transition
	matrix in characteristic modes theory,'' in \emph{2022 {IEEE} International
		Symposium on Antennas and Propagation and {USNC}-{URSI} Radio Science Meeting
		({AP}-S/{URSI})}.\hskip 1em plus 0.5em minus 0.4em\relax {IEEE}, jul 2022.
	
	\bibitem{Maseketal_ModalTrackingBasedOnGroupTheory}
	M.~Masek, M.~Capek, L.~Jelinek, and K.~Schab, ``Modal tracking based on group
	theory,'' \emph{IEEE Trans. Antennas Propag.}, vol.~68, no.~2, pp. 927--937,
	Feb. 2020.
	
	\bibitem{feko}
	\BIBentryALTinterwordspacing
	(2022) Altair {FEKO}. Altair. [Online]. Available:
	\url{https://www.altair.com/feko}
	\BIBentrySTDinterwordspacing
	
	\bibitem{ScatSubstrFEKO}
	\BIBentryALTinterwordspacing
	M.~Capek. (2024) Scattering dyadic characteristic modes (with {A}ltair {FEKO}).
	[Online]. Available:
	\url{https://github.com/kschab/scattering-dyadic-characteristic-modes/tree/main/FEM_MoM-FEKO}
	\BIBentrySTDinterwordspacing
	
	\bibitem{atom}
	\BIBentryALTinterwordspacing
	(2024) {A}ntenna {T}oolbox for {MATLAB} ({AToM}). Czech Technical University in
	Prague. {www.antennatoolbox.com}. [Online]. Available:
	\url{{www.antennatoolbox.com}}
	\BIBentrySTDinterwordspacing
	
	\bibitem{cubeSat}
	\BIBentryALTinterwordspacing
	(2017) Cubesat 101 basic concepts and processes for first-time cubesat
	developers. National Aeronautics and Space Administration (NASA). [Online].
	Available:
	\url{https://www.nasa.gov/wp-content/uploads/2017/03/nasa_csli_cubesat_101_508.pdf}
	\BIBentrySTDinterwordspacing
	
	\bibitem{2021_Deng_AWPL}
	X.~Deng, Y.~Chen, and S.~Yang, ``Characteristic mode formulation for antennas
	with waveguide port feeding structures,'' \emph{IEEE Antennas and Wireless
		Propagation Letters}, vol.~20, no.~10, pp. 2063--2067, 2021.
	
	\bibitem{Kahn_Kurss_1965_MSA}
	W.~Kahn and H.~Kurss, ``Minimum-scattering antennas,'' \emph{IEEE Trans.
		Antennas Propag.}, vol.~13, no.~5, pp. 671--675, 1965.
	
	\bibitem{Rogers_1986_MSA}
	P.~Rogers, ``Application of the minimum scattering antenna theory to mismatched
	antennas,'' \emph{IEEE Trans. Antennas Propag.}, vol.~34, no.~10, pp.
	1223--1228, 1986.
	
	\bibitem{Kim_etal_2013_MSA}
	Y.~G. Kim and S.~Nam, ``Determination of the generalized scattering matrix of
	an antenna from characteristic modes,'' \emph{IEEE Trans. Antennas Propag.},
	vol.~61, no.~9, pp. 4848--4852, 2013.
	
	\bibitem{Alian_2023_MSA}
	M.~Alian and N.~Noori, ``A domain decomposition method for the analysis of
	mutual interactions between antenna and arbitrary scatterer using generalized
	scattering matrix and translation addition theorem of {SWF}s,'' \emph{IEEE
		Trans. Antennas Propag.}, vol.~71, no.~10, p. 8088–8096, Oct. 2023.
	
	\bibitem{Golub+Loan2013}
	G.~H. Golub and C.~F. van Loan, \emph{Matrix Computations}, 4th~ed.\hskip 1em
	plus 0.5em minus 0.4em\relax Baltimore, MD: The Johns Hopkins University
	Press, 2013.
	
\end{thebibliography}
\end{document}